\documentclass[prd,aps,noshowpacs,nofootinbib,superscriptaddress,preprintnumbers
]{revtex4-1}
\usepackage[english]{babel}
\usepackage{amssymb}
\usepackage{amsfonts}
\usepackage{amsmath}
\usepackage{eucal}
\usepackage{graphicx}
\usepackage{epsfig}
\usepackage{mathtools}
\usepackage{color}

\newcommand{\cA}{{\cal A}}  \newcommand{\cB}{{\cal B}}

  \newcommand{\cP}{{\cal P}}

  \newcommand{\cV}{{\cal V}}

\newcommand{\me}{{\mathrm e}} 
\newcommand{\be}{\begin{equation}} \newcommand{\ee}{\end{equation}}
\newcommand{\bea}{\begin{eqnarray}} \newcommand{\eea}{\end{eqnarray}}
\newcommand{\beann}{\begin{eqnarray*}}  \newcommand{\eeann}{\end{eqnarray*}}
\newcommand{\bfig}{\begin{figure}} \newcommand{\efig}{\end{figure}}
\newcommand{\ba}{\begin{array}} \newcommand{\ea}{\end{array}}
\newcommand{\bcen}{\begin{center}} \newcommand{\ecen}{\end{center}}
\newcommand{\btab}{\begin{tabular}} \newcommand{\etab}{\end{tabular}}

\def\tr{\operatorname{tr\:}}     
   
     \def\diag{\operatorname{diag}}
     \def\sign{\operatorname{sign}}
\renewcommand{\Re}{\mathop{\rm Re}}   \renewcommand{\Im}{\mathop{\rm Im}}
\newcommand{\dd}{{\rm d}}
\newcommand{\e}{{\rm e}}

\newtheorem{Proposition}{Proposition}[section]

\newtheorem{Theorem}{Theorem}[section]
\newtheorem{Lemma}{Lemma}[section]
\newtheorem{Corrolary}{Corrolary}[section]

\newcommand{\bp}{\begin{Proposition}}	\newcommand{\ep}{\end{Proposition}}
\newcommand{\bt}{\begin{Theorem}}	\newcommand{\et}{\end{Theorem}}
\newcommand{\bl}{\begin{Lemma}}		\newcommand{\el}{\end{Lemma}}
\newcommand{\bc}{\begin{Corrolary}}	\newcommand{\ec}{\end{Corrolary}}
\begin{document}

\title{Frequency dependence of the Chiral Vortical Effect}

\author{Karl Landsteiner}
\email{karl.landsteiner@csic.es}
\affiliation{Instituto de F{\'\i}sica Te\'orica, IFT-UAM/CSIC, Universidad
Aut\'onoma de Madrid, Cantoblanco E-28049 Madrid, Spain}

\author{Eugenio Meg\'{\i}as}
\email{emegias@ifae.es}
\affiliation{Grup de F\'{\i}sica Te\`orica and IFAE, Departament de F\'{\i}sica,
Universitat Aut\`onoma de Barcelona, Bellaterra E-08193 Barcelona, Spain}

\author{Francisco \surname{Pe\~na-Ben\'{\i}tez}}
\email{pena@physics.uoc.gr}
\affiliation{Instituto de F{\'\i}sica Te\'orica, IFT-UAM/CSIC, Universidad
Aut\'onoma de Madrid, Cantoblanco E-28049 Madrid, Spain}
\affiliation{Departamento de F\'isica Te\'orica, Universidad Aut\'onoma de
Madrid, Cantoblanco E-28049 Madrid, Spain}

\affiliation{Crete Center for Theoretical Physics, Department of Physics,
University of Crete, 71003 Heraklion, Greece}

\begin{abstract}
We study the frequency dependence of all the chiral vortical and
magnetic conductivities for a relativistic gas of free chiral fermions
and for a strongly coupled conformal field theory with holographic dual in four
dimensions. Both systems have gauge and gravitational anomalies, and
we compute their contribution to the conductivities.  The chiral
vortical conductivities and the chiral magnetic conductivity in the
energy current show a 
frequency dependence in the form of a delta centered at zero frequency.  This highly discontinuous
behavior is a natural consequence of the Ward identities that include the energy momentum tensor.
We discuss the physical interpretation of
this result and its possible implications for the quark gluon plasma
as created in heavy ion collisions.
In the Appendix we discuss why the chiral magnetic effect seems to
vanish in the {\em consistent} current for a particular implementation
of the axial chemical potential.
\end{abstract}

\preprint{IFT-UAM/CSIC-13-132, UAB-FT-748,  CCTP-2013-22, CCQCN-2013-12}
\maketitle
%
\section{Introduction} 
\label{sec:intro}

During the last years the study of anomaly induced transport coefficients 
has proved a subject of increasing interest. 
Charge separation found in RHIC and confirmed more 
recently at the LHC \cite{Abelev:2009uh,Abelev:2012pa} can possibly
be traced back to the  chiral magnetic
Effect (CME). This effect says that a system with triangle 
anomalies in an external magnetic field will show an electric current 
parallel to the magnetic field \cite{Fukushima:2012vr}
\begin{equation}
 \vec J = \frac{e \mu_5}{2\pi^2}e\vec B. 
\end{equation}
There have been early precursors that studied manifestation of this phenomena in
neutrino physics~\cite{Vilenkin:1995um,Vilenkin:1980fu}, the early
universe~\cite{Giovannini:1997eg} and condensed matter
systems~\cite{Alekseev:1998ds}. In recent years the increasing
interest in this effect has been spurred by the role it might play in 
the physics of the quark gluon plasma.

But this phenomenon is not the only one present in a chiral system at
finite temperature and/or chemical potential. The presence of a
magnetic field can also produce an axial current known as the chiral
separation effect (CSE)
\cite{Son:2004tq,Metlitski:2005pr,Newman:2005as} and a vortex can
contribute to the electric and axial current, this is the so-called
chiral vortical effect (CVE)
\cite{Kharzeev:2007tn,Erdmenger:2008rm,Banerjee:2008th,Son:2009tf}.
Apart from charge flow in a relativistic fluid there exists also
energy flow and analogous anomaly related transport
effects in the energy current
$ J^i_\epsilon =T^{0i}$. In this paper we will use the compact notation
$J^i_A\in\{J^i_e,J^i_5,J^i_\epsilon\}$, where we include the electric,
axial and energy currents. With this notation we can write two compact
Kubo formulae for the chiral magnetic and vortical conductivities
\cite{Kharzeev:2009pj,Amado:2011zx,Landsteiner:2012kd}.
\begin{eqnarray}
 \label{eq:Kubo1} \sigma^\cB_A(\omega,\vec k) = -
 \sum_{i,j}\epsilon_{ijn}\frac{i}{2k_n}\langle J^i_A   J^j_e \rangle  \,,\\
 \label{eq:Kubo2} \sigma^\cV_A(\omega,\vec k) = -
 \sum_{i,j}\epsilon_{ijn}\frac{i}{2k_n}\langle J^i_A   J^j_\epsilon \rangle\,.
\end{eqnarray} 
The most significant result of anomalies is that they produce equilibrium 
currents. These equilibrium conductivities are defined via the Kubo formulae
in the kinematic region in which first the frequency is set to zero  and then 
the limit to zero momentum is taken. 

As the quark gluon plasma produced in a heavy ion collision has a finite
life time and size, it is mandatory to know the full frequency and
momentum dependence of the response to magnetic field and vorticity.  A
detailed study of the frequency dependence of the chiral magnetic
effect at weak coupling was done in \cite{Kharzeev:2009pj} and in a
strongly coupled regime using holography in \cite{ Yee:2009vw}\footnote{In 
a condensed matter context the frequency dependence and in particular the
non-analytic behavior under exchange of the limits $\omega\rightarrow 0$ and
$k\rightarrow 0$ of the CME was also emphasized in \cite{ChenWuBurkov}.}.  In
the study of the chiral vortical conductivity in a static situation
using Kubo formulae at weak coupling a surprising result was found
\cite{Landsteiner:2011cp}. A purely temperature dependent term
appeared in the conductivity consistent with previous
hydrodynamical analysis\footnote{Using hydrodynamical considerations a
  term with the same form was found, but the numerical coefficient
  multiplying the temperature was completely undetermined by the
  method \cite{Neiman:2010zi}. This contribution had also been found earlier in
  \cite{Vilenkin:1980zv}.}, but it was realized that this contribution
is present if and only if the theory has a mixed gauge-gravitational
anomaly. To verify that result at strong coupling a bottom up
holographic model was built introducing a mixed gauge-gravitational
anomaly into the system and the same contribution appeared in this
holographic setup \cite{Landsteiner:2011iq}. Actually this result has
been confirmed many times using different
approaches~\cite{Loganayagam:2012pz,Chapman:2012my,Gao:2012ix,Jensen:2012kj,
Megias:2013joa,
Megias:2013xla,Megias:2013uua}. Anomalous
conductivities are therefore sensitive to both, pure gauge
and mixed gauge-gravitational anomalies.  It is understood by now that in
theories in which the anomaly is purely classical, e.g. neither the
gauge fields nor the metrics are considered quantum variables, the
anomalous equilibrium transport is subject to a non-renormalization
theorem \cite{Son:2009tf, Jensen:2012kj, Golkar:2012kb, Gorbar:2013upa,
 Hou:2012xg,Zakharov:2012vv, Jensen:2013vta}  .

A computation of the frequency dependence of the CVE was done in
\cite{Amado:2011zx} within a holographic model with a pure gauge anomaly only. 
However the contribution of the gravitational anomaly could be a 
leading term in heavy ion
collisions where the temperature reached is much higher than the
chemical potential. Therefore it is necessary to consider both anomalies. In the
experiment the charge separation due to the CME and CVE in the vector current 
should be seen through the search of charged particles in the perpendicular
directions to the reaction plane \cite{Kharzeev:2010gr}. And the signature left
by the
separation of
chirality is predicted to be an enhanced production of higher spin
mesons after the freezeout~\cite{KerenZur:2010zw}.

As we previously said a more realistic analysis of the CVE is needed.
We take this as the motivation to compute the frequency and momentum dependence 
of the chiral vortical conductivity  in the electric, axial and energy currents 
at weak and strong coupling. 

At weak coupling this implies working out the sum over Matsubara frequencies.
We take this as an opportunity to give a careful discussion of the 
seeming ``gauge''
dependence on the result for the CME in the Appendix. It is well established 
\cite{Rebhan:2009vc, Landsteiner:2012kd,Jensen:2012kj}
that the CME 
receives an additional contribution depending on the axial gauge potential if
formulated
in terms of the {\em consistent} current. We show how this appears at weak
coupling and
we give a physical interpretation to the different responses in {\em consistent}
and {\em covariant} currents. 

The manuscript is organized as follows.  In
section~\ref{sec:weak_coupling} we consider a gas of free fermions
with a $U(1)_V\times U(1)_A$ global symmetry and compute the frequency
and momentum dependence of all the anomaly induced transport
coefficients using Kubo formulae.  We present in
section~\ref{sec:strongly_coupled_regime} a numerical computation of
the conductivities using a holographic model describing a strongly
coupled plasma of fermions with the same symmetry group as in the
weakly coupled case. In section~\ref{sec:hydro} we compute the
anomalous transport predicted by hydrodynamics, and compare with the
strong coupling results. Finally we discuss the role of the CVE in
heavy ion collisions and draw our conclusions in
section~\ref{sec:discussions}.  In the Appendix we discuss the
subtleties arising in the sum over Matsubara frequencies when dealing
with chemical potentials for anomalous symmetries.

\section{Weakly coupled regime}
\label{sec:weak_coupling}

We define the chemical potential through boundary conditions on the
fermion fields around the thermal circle \cite{Landsman:1986uw}, $
\Psi^f(\tau) = - e^{\beta \mu^f} \Psi^f(\tau-\beta)$ with $\beta=1/T$.
Therefore the eigenvalues of $\partial_\tau$ are
$i\tilde\omega_n+\mu^f$ for the fermion species $f$ with
$\tilde\omega_n=\pi T(2n+1)$ the fermionic Matsubara frequencies.
From now on we will consider the symmetry group $U_V(1)\times U_A(1)$,
i.e. one vector and one axial current with chemical potentials
$\mu_{\pm} = \mu \pm \mu_5$, charges $q^+_{v,5} = (1,1)$ and
$q^-_{v,5} = (1,-1)$ for one right-handed and one left-handed
fermion. A convenient way of expressing the currents is in terms of
Dirac fermions and writing
\begin{eqnarray}
J^i_{e,5}(x) &=& \bar\Psi(x) \gamma^i  Q_{e,5}\Psi(x) \,, \label{eq:Jv5}\\
J^{i}_\epsilon(x) &=&  \frac i 4 \bar\Psi(x) ( \gamma^0 \overleftrightarrow
{\partial^i} + \gamma^i  \overleftrightarrow{\partial^0}  ) \Psi(x) \,,
\label{eq:JE}
\end{eqnarray}
where the vector charge is $Q_e = {\mathcal P}_{+} + {\mathcal P}_{-}$ and the
axial charge is $Q_5=\mathcal P_+-\mathcal P_-$. $J_e$, $J_5$ and $J_\epsilon$
correspond to the vector, axial and energy currents, respectively.
We used the chiral projector $\cP_\pm = \frac 1 2 (1\pm\gamma_5)$.  Our metric
is $g_{\mu\nu} = \diag(1,-1,-1,-1)$. The fermion propagator is
\begin{eqnarray}
S(q) &=&  \frac{1}{2} \sum_{s,t=\pm} \Delta_t(i\tilde\omega_s,\vec{q}) \cP_s
\gamma_\mu \hat q^\mu_t \,,   \label{eq:Sq}\\
\Delta_t( i\tilde\omega_s, q) &=& \frac{1}{i\tilde\omega_s - t E_q}\,,
\label{eq:Delta_t}
\end{eqnarray}
with  $i\tilde\omega_s = i\tilde\omega_n + \mu_s$, $\hat q_t^\mu = (1, t \hat
q)$, $\hat{q} = \frac{\vec{q}}{E_q}$. We will consider massless
fermions, so that $E_q=|\vec q |$.  The value $t = +1$ corresponds to particles
(positive energy) and $t = -1$ to antiparticles (negative energy). Label $s$
refers to right-handed ($s=+1$) and left-handed ($s=-1$) chiralities, so that
right and left chemical potentials are related to baryon and axial chemical
potentials as $\mu_s = \mu + s \mu_5$.

\subsection{Chiral vortical conductivities}
\label{subsec:vortical-conductivity}

Since we have the Kubo formulae, the problem of computing the
transport coefficients, Eqs.~(\ref{eq:Kubo1}) and (\ref{eq:Kubo2}),
reduces to the computation of the retarded correlator between the
currents $J_A^i$
\begin{equation}
  \label{eq:retarded}G_{AB}(x-x^\prime) = \frac{1}{2} \epsilon_{ijn}\,i \,
\theta(t-t^\prime)  \,\langle [J^i_{A}(x),J^{j}_B(x^\prime)] \rangle \,, 
\end{equation}
in particular we will focus on the case of the vortical conductivity
in which the second current in the formula~(\ref{eq:retarded}) is the
energy flux $J^i_\epsilon$. The generalization to the magnetic case is
straightforward, and we will address it in
Sec.~\ref{subsec:magnetic-conductivity}. Let us redefine the
correlators associated with the chiral vortical effect as
\begin{equation}
  G_A^\cV \equiv G_{A\epsilon} \,, \qquad A = e, 5, \epsilon \,.
\end{equation}
The one loop correlators $G_5^\cV$ and $G_\epsilon^\cV$ can be computed,
respectively, as
\begin{center}
\begin{figure}[tbp]
\includegraphics{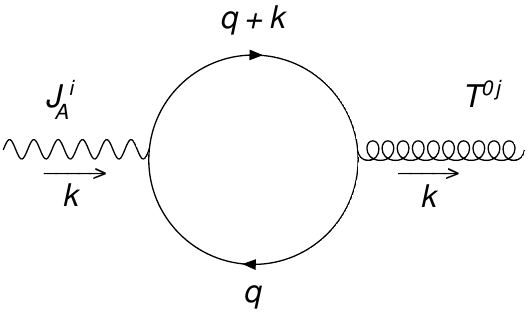}
\caption{One loop diagram contributing to the chiral vortical conductivities
$G_A^\cV$, see 
Eqs.~(\ref{eq:Gav}) and (\ref{eq:Gaepsilon}). For $A=\epsilon$ there is also the
contribution coming 
from the seagull diagram, see~\cite{Manes:2012hf}.}
\label{fig:1loop}
\end{figure}
\end{center}
\begin{eqnarray}
G_5^\cV(k_0,k) &=& \frac{1}{8\beta}\sum_{\tilde\omega} \int\frac{d^3q}{(2\pi)^3}
\epsilon_{ijn} \tr \Bigg[ S(q) \gamma^i \gamma_5 S(q+k) \left(  \gamma^0 (2q^j +
k^j) + \gamma^j (2 i\tilde\omega + k^0 )\right)  \Bigg] \,, \label{eq:Gav}
\end{eqnarray}
\begin{equation}
G_\epsilon^\cV(k_0,k) = \frac{1}{16\beta}\sum_{\tilde\omega}
\int\frac{d^3q}{(2\pi)^3}
\epsilon_{ijn} \tr \Bigg[ S(q) \gamma^i S(q+k) \left( 2 \gamma^0 (2q^j +
k^j) + \gamma^j (2 i\tilde\omega + k^0 )  \right) (2 i\tilde\omega + k^0 ) 
+ S(q) \left\{ \gamma^i \gamma^j , k\!\!\!\!/ \right\} \Bigg] \,,
\label{eq:Gaepsilon}
\end{equation}
where $k\!\!\!\!/ = \gamma^\mu k_\mu = \gamma^0 k^0 - \gamma^m
k^m$. Figure~\ref{fig:1loop} shows the one loop diagram
corresponding to $G_A^\cV$. The expression for $G_e^\cV(k_0,k)$ is the
same as Eq.~(\ref{eq:Gav}) but removing the~$\gamma_5$ matrix in the
integrand. The last term inside the bracket in
Eq.~(\ref{eq:Gaepsilon}) corresponds to the seagull diagram which was computed
in~\cite{Manes:2012hf}. 
The correlators $G_e^\cV(k_0,k)$
and $G_5^\cV(k_0,k)$ have been computed in detail in
Ref.~\cite{Landsteiner:2011cp} at zero frequency, and the computation of
$G_\epsilon^\cV(k_0,k)$ follows straightforwardly by using the same
procedure, so we will skip here the technical details. An evaluation of
Eqs.~(\ref{eq:Gav}) and (\ref{eq:Gaepsilon}) leads to the result
(from now on we denote $q=|\vec{q}|$ and $k=|\vec{k}|$)
\begin{equation}
\widehat{G}_{A}^\cV(k_0,k) = \frac{ik_n}{16\pi^2}\int_0^\infty dq  \,
f_A^\cV(q)g_A^\cV(q)\,, \label{eq:GcV}
\end{equation}
where
\begin{equation}
 g_A^\cV(q)=\left\{
\begin{array}{ll}
     q   \left(1 - 3\frac{k_0^2}{k^2}\right) + \frac{(k^2-k_0^2)}{8
k^3}\sum_{t=\pm}\left[3 k_0^2 - k^2 +12q(q + k_0 t)\right] \log \left[
\frac{\Omega_t^2 - (q+k)^2}{\Omega_t^2 - (q-k)^2}\right]       \quad           &
 A = e,5 \\
          2 q^2 \left(1 - 2 \frac{k_0^2}{k^2} \right)  +   \frac{(k^2 -
k_0^2)}{16 k^3}\sum_{t=\pm}(2 q + k_0 t)( 2 k_0^2 - k^2  + 8q(q + k_0 t) ) \log
\left[
\frac{\Omega_t^2 - (q+k)^2}{\Omega_t^2 - (q-k)^2}\right]      \quad           & 
A = \epsilon
    \end{array}
    \right. \label{eq:gAV}
\end{equation}
with $\Omega_t = k_0 + i\epsilon + t q$, and
\begin{equation}
 f_A^\cV(q)=\sum_{s,t=\pm} s^\alpha t^\gamma  n(E_q - t\mu_s) \,,\qquad 
\left\{\begin{array}{l}
        \alpha=1,\gamma=0 \quad, \quad A=e    \\
      \alpha=0,\gamma=0 \quad, \quad A=5                \\
       \alpha=1,\gamma=1 \quad , \quad A=\epsilon      
    \end{array}
    \right. \,, \label{eq:fAV}
\end{equation}
where $n(x) = 1/(e^{\beta x} + 1)$ is the Fermi-Dirac distribution
function. The hat in Eq.~(\ref{eq:GcV}) denotes the vacuum subtracted
contribution (see the Appendix). To compute the
imaginary part of Eqs.~(\ref{eq:GcV}) and (ref{eq:gAV}) in the same
spirit of \cite{Kharzeev:2009pj} we need the relations
\begin{align}
 \Im\sum_{t=\pm} \log\left[ 
\frac{(k_0+i\epsilon+t q)^2-(k+q)^2}{(k_0+i\epsilon+t q)^2-(k-q)^2} \right] =&
\pi\, \left[ \theta(k_+ - q) - \theta(k_--q) \right]\,,\\
\nonumber \Im\sum_{t=\pm} t \log\left[ 
\frac{(k_0+i\epsilon+t q)^2-(k+q)^2}{(k_0+i\epsilon+t q)^2-(k-q)^2} \right] =&
-\pi\,\theta(k_0^2-k^2)\, \mathrm{sgn}(k_0) \,\left[ \theta(k_+ - q) -
\theta(k_--q) \right] + \,\\
 &+ \pi\, \theta(k^2-k_0^2)\,\left[ \theta(q-k_+) + \theta(q-k_-) \right] \, ,
\end{align}
where $\theta(x)$ is the step function and $k_\pm =\frac{1}{2}|k_0\pm k|$. From
an analytical evaluation of Eqs.~(\ref{eq:GcV}) and (\ref{eq:gAV})
for $A=e,5,\epsilon$ one gets the following momentum and frequency
dependence of the vector, axial and energy vortical conductivities,
\begin{eqnarray}
\Im[\sigma^\cV_A(k_0,k)] &=&\frac{1}{16\pi}\frac{k^2-k_0^2}{4k^3}\bigg\{
2k_0\theta(k^2-k_0^2)\left[24\mu\mu_5\delta_{A,e} +
\left(12\left(\mu^2+\mu_5^2+\frac{\pi^2}{3}T^2\right) +
(k_0^2-k^2)\right)\delta_{A,5} \right]\nonumber\\
&&-T\sum_{r,s,t=\pm}r s^\alpha \bigg[  k^2
\log{\left[1+\me^{\beta(k_r+t\mu_s)}\right]} -12T^2  \text{Li}_3\left(-e^{\beta 
(k_r+t \mu_ s)}\right) \nonumber\\
&&+6kT\mathrm{Li}_2\left(-\me^{\beta(k_r+t\mu_s)}
\right)\left(\theta(k^2-k_0^2)+r\sign(k_0)\theta(k_0^2-k^2) \right)
\bigg]\bigg\} \,, \qquad A= e, 5 \,,
\end{eqnarray}
\begin{eqnarray}
\Im[\sigma^\cV_\epsilon(k_0,k)] &=&\frac{1}{16\pi}\frac{k^2-k_0^2}{16k^3}\bigg\{
8k_0\mu_5\theta(k^2-k_0^2)\left[8(3\mu^2+\mu_5^2) + \left( 2k_0^2-k^2 + 2(2\pi
T)^2\right) \right]+\nonumber\\
&&\sum_{r,s,t=\pm}r s t \bigg[ 10k^2T^2  \text{Li}_2\left(-e^{\beta  (k_r+t \mu_
s)}\right) +96T^4  \text{Li}_4\left(-e^{\beta  (k_r+t \mu_ s)}\right)\\
&&+kT\left(
k^2\log{\left[1+\me^{\beta(k_r+t\mu_s)}\right]}-48T^2\mathrm{Li}_3\left(-\me^{
\beta(k_r+t\mu_s)}\right) \right)\left(
\theta(k^2-k_0^2)+r\sign(k_0)\theta(k_0^2-k^2) \right) \bigg]\bigg\} \,,\nonumber
\end{eqnarray}
where $\text{Li}_n(x)$ is the polylogarithm function of order $n$. A series
expansion at small $k$ of these expressions leads to
\begin{eqnarray}
\Im[\sigma^\cV_A(k_0,k)] &=& 
\theta(k^2-k_0^2) \frac{1}{16\pi}\frac{k^2-k_0^2}{4k^3}
k_0\bigg[24\mu\mu_5\delta_{A,e} +
\left[12\left(\mu^2+\mu_5^2+\frac{\pi^2}{3}T^2\right) +
(k_0^2-k^2)\right]\delta_{A,5} \nonumber \\
&&+ \mu_5\left[8(3\mu^2+\mu_5^2+ \pi^2 T^2) + \left( 2k_0^2-k^2 \right) \right]
\delta_{A,\epsilon}\bigg]\nonumber\\
&&-k^2\theta(k_0^2-k^2) \frac{k_0|k_0|}{2^{10}\times15\pi}  \sum_{s,t=\pm}
s^\alpha t^\gamma \left(\frac{d}{dx}\right)^{2-\gamma} n(2x-t \mu_s)\Bigg|_{x=
\frac{1}{4}|k_0|} + {\cal O}(k^4) \,.
\end{eqnarray}
Notice that an expansion at small $k$ in the term $\sim\theta(k^2-k_0^2)$
demands that one considers $|k_0| \lesssim k$, otherwise this contribution is
vanishing. This restriction does not apply in the term $\sim\theta(k_0^2-k^2)$. 
In the limit $k\to 0$ this expression leads to the result
\begin{equation}
\Im [ \sigma_A^\cV(\omega,0)] = \pi \sigma_{A,(0)}^\cV \omega\delta(\omega) \,,
\qquad \sigma_{A,(0)}^\cV = \left\{
\begin{array}{l}
 \frac{1}{2\pi^2} \mu\mu_5    \quad           \\ 
 \frac{1}{4\pi^2}  \left( \mu^2 + \mu_5^2 + \frac{\pi^2}{3} T^2 \right)   \\
 \frac{\mu_5}{6\pi^2} \left( 3\mu^2 + \mu_5^2 + \pi^2 T^2 \right)    
 \end{array}
 \right.\,, \label{eq:sigmaImV}
\end{equation}
where we have made use of the fact that $\lim_{k\to 0}
\theta(k^2-k_0^2)\frac{k^2-k_0^2}{k^3} = \frac{4}{3}\delta(k_0)$. In
Eq.~(\ref{eq:sigmaImV}) we have denoted the frequency as $\omega$. In the
following we will use either $k_0$ or $\omega$. Using the Kramers-Kronig
relation one can obtain the real part of the conductivities at $k=0$ and
$\omega$ finite, and they read
\begin{eqnarray}
&&\Re[\sigma_A^\cV(\omega,0)] = \left\{ 
    \begin{array}{ll}
       \sigma_{A,(0)}^\cV   & \quad  \omega = 0 \\
        0                 &  \quad \omega \ne  0 
    \end{array}
    \right. \,.
\end{eqnarray}
It is remarkable that the chiral vortical conductivities in the free
field theory are zero at finite frequency and zero momentum. 
The discontinuous behavior at $\omega=0$ is also of great relevance. 
We show in fig.~\ref{fig:3Dvorticale} the full frequency
and momentum dependence of $\Re[\sigma^\cV_e(\omega,k)]$ at low and
high temperatures. We have introduced the dimensionless parameter $\tau
= 2\pi T/\mu$ in order to have a better comparison with the results
from holography in Sec.~\ref{sec:strongly_coupled_regime},
variables. The figures have three features: i) at high temperature, there is a peak 
at $\omega = k$, ii) at low temperature, in 
addition to the peak at $\omega = k$, there are peaks at $\omega = k
\pm 2\mu_s$, iii) the conductivities are vanishing 
at $k=0$, $\omega\ne 0$, and they present a discontinuity at $k=0$, $\omega=0$.
From their behavior and these features 
one can see that the vortical conductivities are approximately vanishing  at
high temperature, in the regime $\omega > k$, see fig.~\ref{fig:3Dvorticale}. 
We will confront these results with the ones predicted at strong coupling in
Sec.~\ref{sec:strongly_coupled_regime}, 
and discuss their implications in Sec.~\ref{sec:discussions}.

\begin{center}
\begin{figure}[tbp]
\includegraphics[angle=0,height=0.3\textwidth]{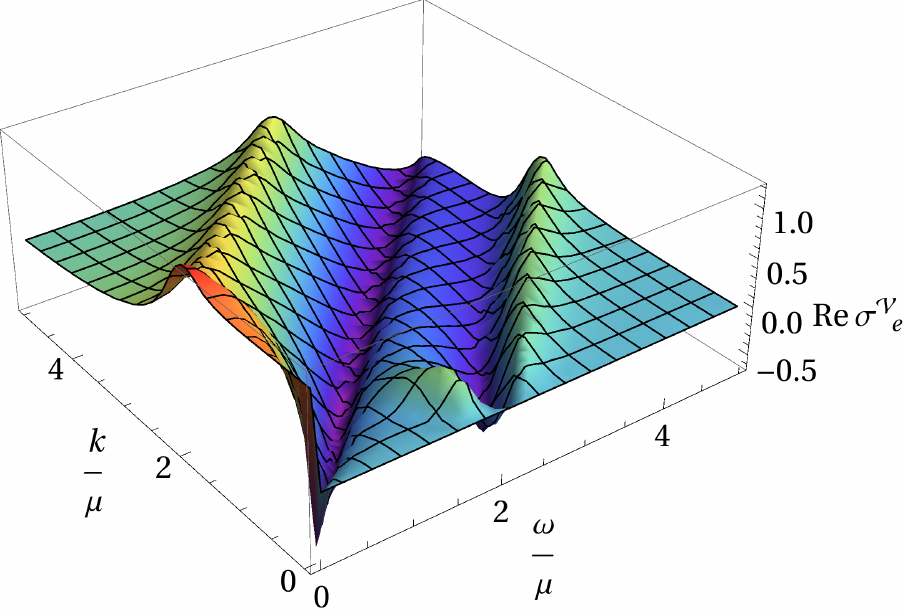} 
\hspace{1cm}
\includegraphics[angle=0,height=0.3\textwidth]{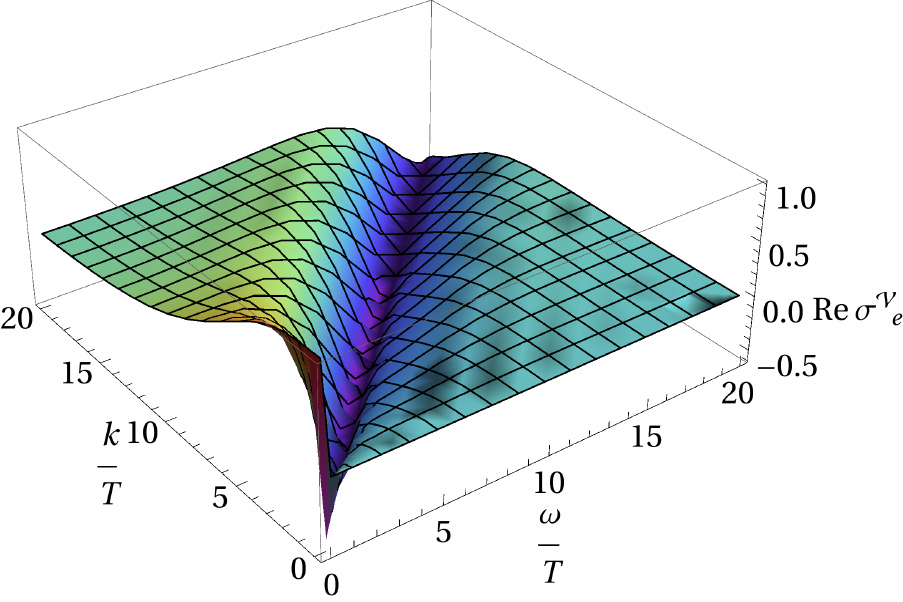}
\caption{Frequency and momentum dependence of the real part of the
  vector vortical conductivity~$\sigma_e^\cV(\omega,k)$ normalized to
  its zero frequency-momentum value, from a numerical evaluation of
  Eqs.~(\ref{eq:GcV}) and (\ref{eq:fAV}). We consider  $\mu = 10 \,\textrm{MeV}$,
$\mu_5 = 1 \,\textrm{MeV}$, 
  and temperature below the QCD phase transition $\tau = 0.24$ (left figure) and
above the phase transition $\tau = 440$ (right figure). [Color online]}
\label{fig:3Dvorticale}
\end{figure}
\end{center}

\subsection{Chiral magnetic conductivity and chiral separation effect}
\label{subsec:magnetic-conductivity}

We will extend for completeness the computation of the finite
frequency and momentum behavior of conductivities to the chiral
magnetic and separation effects. The chiral magnetic conductivity was
studied in~\cite{Kharzeev:2009pj} in the case of a $U(1)_V\times
U(1)_A$ symmetry at weak coupling. For a general symmetry group all
the DC magnetic conductivities were computed
in~\cite{Landsteiner:2011cp}. They follow from the retarded Green's
function of two charge currents
\begin{equation}
G_A^\cB(x-x^\prime) = \frac{1}{2} \epsilon_{ijn}\,i \, \theta(t-t^\prime) 
\,\langle [J^i_A(x),J_{e}^j(x^\prime)] \rangle \,, \qquad A = e, 5, \epsilon \,.
\label{eq:GB}
\end{equation}
Following a similar procedure as in the chiral vortical computation of
Sec.~\ref{subsec:vortical-conductivity}, we get that the retarded correlator can
be written as
\begin{eqnarray}
\widehat G^{\cB}_A(k_0,k) &=&
\frac{i}{16\pi^2}\frac{k_n}{k}\frac{k^2-k_0^2}{k^2} \int_0^\infty dq \,
f_A^\cB(q) \sum_{t=\pm} (2q + k_0 t)  \log\left[\frac{\Omega_t^2 -
(q+k)^2}{\Omega_t^2 - (q-k)^2}\right] \,,  \label{eq:GBA}
\end{eqnarray}
where 
\begin{equation}
 f_A^\cB(q)=\sum_{s,t=\pm} s^\alpha t \, n(E_q-t\mu_s) \,, \qquad A = e, 5 \,,  
\label{eq:fBA}
\end{equation}
and the value of $\alpha$ is defined as in Eq.~(\ref{eq:fAV}). We do not show in
Eqs.~(\ref{eq:GBA}) and (\ref{eq:fBA}) explicit formulas for
$G_\epsilon^\cB(k_0,k)$, as in the free field theory they are
identical to $G_e^\cV(k_0,k)$, which was presented in
Eqs.~(\ref{eq:GcV}) and (ref{eq:fAV}).  This can easily be checked from
the structure of the correlation functions, c.f.~Eq.~(\ref{eq:Gav}).

\begin{center}
\begin{figure}[tbp]
\includegraphics[angle=0,width=0.49\textwidth]{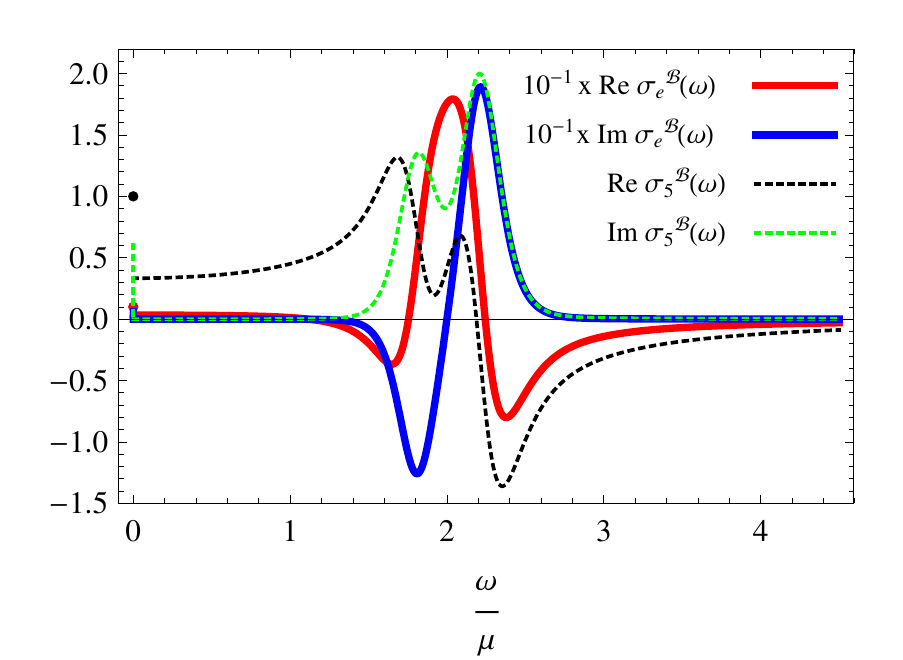}
\includegraphics[angle=0,width=0.499\textwidth]{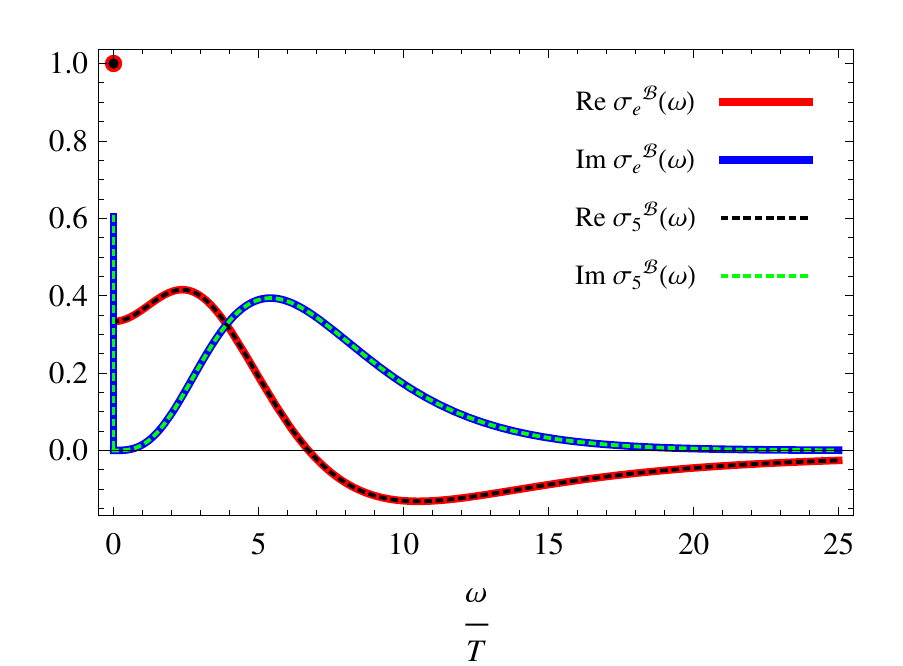} 
\caption{Frequency dependence of the vector and axial magnetic
  conductivities, $\sigma_e^\cB$ and $\sigma_5^\cB$, normalized to
  their zero frequency value, from a numerical evaluation of
  Eqs.~(\ref{eq:GBA}) and (\ref{eq:fBA}).  We consider $k=0$, $\mu = 10
  \,\textrm{MeV}$, $\mu_5 = 1 \, \textrm{MeV}$, $\tau = 0.24$ (left
  figure) and $\tau = 440$ (right figure). The vertical lines at $\omega=0$ 
  in the imaginary parts are meant to remind one of the presence of the
  term $\sim\omega \delta(\omega)$. [Color online]}
\label{fig:wcchiralB}
\end{figure}
\end{center}

The frequency dependence  for $\sigma^\cB_e$ was originally computed
in~\cite{Kharzeev:2009pj}. Here we provide analytical results for this and other
conductivities. In a series expansion at small $k$, the imaginary part writes
\begin{eqnarray}
\Im[\sigma^\cB_A(k_0,k)] &=& \theta(k^2-k_0^2) \frac{k^2-k_0^2}{4\pi k^3}k_0 
\left[\mu_5 \delta_{A,e} + \mu \delta_{A,5} + \frac{3}{2}\mu\mu_5
\delta_{A,\epsilon}\right]  \\
&&+ \theta(k_0^2-k^2) \frac{k_0|k_0|}{96\pi} \sum_{s,t=\pm} t\zeta_A \left[
\frac{d}{dx} + k^2\left(\frac{1}{k_0^2}\frac{d}{dx} +
\frac{1}{40}\frac{d^3}{dx^3} \right) \right]n(x+t\mu_s) \bigg|_{x = |k_0|/2}  +
{\cal O}(k^4) \,,
 \quad \zeta_A =   \left\{ 
    \begin{array}{l}
        s        \\
        1        \\
        0      
    \end{array}
    \right. \,. \nonumber
\end{eqnarray}
In the limit $k\to 0$ this expression leads to
\begin{equation}
\Im [\sigma^\cB_A(\omega,0)] =  \left(2 + \delta_{A,\epsilon} \right)
\frac{\pi}{3} \sigma^\cB_{A,(0)} \omega\delta(\omega) +
\frac{\omega|\omega|}{96\pi} \sum_{s,t=\pm} t \zeta_A \left[\frac{d}{dx} n(x+t
\mu_s) \right]_{x = |\omega|/2} \,,
  \label{eq:ImsigmaB}
\end{equation}
where $\sigma^\cB_{A,(0)}$ are given by
\begin{equation}
\sigma^\cB_{A,(0)} = \frac{1}{2\pi^2}  \left\{
\begin{array}{l }
      \mu_5     \quad            \\
       \mu     \quad              \\
       \mu\mu_5  \quad        
    \end{array}
    \right.  \,.  \label{eq:sigmaB_0}
\end{equation}
Finally in the zero temperature limit, the imaginary part of these
conductivities becomes
\begin{equation}
\Im [ \sigma^\cB_A (\omega,0)] =  \left(2 +
\delta_{A,\epsilon}\right)\frac{\pi}{3}\sigma^\cB_{A,(0)}\omega\delta(\omega) -
\frac{\omega |\omega|}{96\pi} \sum_{s,t=\pm} t \zeta_A \delta(\omega/2 + t\mu_s)
\,. \label{eq:ImsigmaBT0}
\end{equation}
The real part can be recovered by using the Kramers-Kronig relation, and it
reads
\begin{equation}
\Re [\sigma^\cB_A (\omega,0)] =  \left\{ 
    \begin{array}{ll}
       \sigma^\cB_{A,(0)}   \qquad     &    \omega = 0 \\
        \frac{1}{3\pi^2} \sum_{s=\pm} \zeta_A\frac{\mu_s}{4-(\omega/\mu_s)^2} 
\qquad &    \omega \ne 0 
    \end{array}
    \right. \,.
   \label{eq:ResigmaB} 
\end{equation}
Again using the Kubo formulae, Eq.~(\ref{eq:Kubo1}), and evaluating
the conductivities at zero frequency we get the DC magnetic
conductivities. Even at $k=0$ it is not obvious how to find an analytic
expression for the real part of the conductivities at finite temperature. So we
plot in fig.~\ref{fig:wcchiralB} the frequency dependence of the
chiral magnetic $\sigma^\cB_e$ and chiral separation $\sigma^\cB_{5}$
conductivities respectively, for different values of temperature. Our
result for $\sigma^\cB_e(\omega,0)$ agrees with the one obtained
in~\cite{Kharzeev:2009pj}. At low temperature one can identify in
fig.~\ref{fig:wcchiralB} (left) the two resonances in $\omega = 2\mu_+ \,,
2 \mu_-$ obtained in Eq.~(\ref{eq:ImsigmaBT0}). Note that both resonances
have the same sign in $\Im\sigma^\cB_5$, and opposite signs in
$\Im\sigma^\cB_e$. When temperature increases, the delta functions
are smoothed out. One can easily evaluate from Eq.~(\ref{eq:ImsigmaB})
that the width of the resonances increases with temperature linearly. It is worth mentioning that the peaks in $\omega = k \pm 2\mu_{s}$, which appear in the chiral vortical and chiral magnetic conductivities at low temperatures, see e.g. fig.~\ref{fig:3Dvorticale} (left), become a delta function when $T\to 0$ only in the case $k=0$ for the chiral magnetic conductivities, so these peaks have a resonant character only in this case. In the chiral vortical conductivities these peaks disappear when $k=0$, as it can be seen in fig.~\ref{fig:3Dvorticale} (left).

At very high temperatures these two peaks disappear, and in this case
the position of the single peak appearing in
figure~\ref{fig:wcchiralB} (right) is not related to the value of
chemical potentials, but it depends linearly on temperature. It was
already shown in~\cite{Kharzeev:2009pj} that $\Im\sigma^\cB_e$ at high
temperature has a single peak at $\omega \simeq 5.406 \, T$, and these
authors derived a simple formula in this regime by expanding
Eq.~(\ref{eq:ImsigmaB}) for $T/\mu_s \gg 1$. We have seen that the
same formula applies for $\sigma^\cB_5$ at leading order in this
expansion (after the replacement $\mu_5 \to \mu$), i.e.
\begin{equation}
\Im [\sigma^\cB_A(\omega,0)] \approx  (2 + \delta_{A,\epsilon})\frac{\pi}{3} \sigma^\cB_{A,(0)} \omega\delta(\omega) + \frac{\omega|\omega|}{24\pi T^2 } \left( n(|\omega|/2) \right)^3 \left( e^{|\omega|/T} - e^{|\omega|/(2T)} \right) \left(\mu_5 \delta_{A,e} + \mu \delta_{A,5} \right) \,, \qquad T/\mu_s \gg 1 \,, \label{eq:ImsigmaT}
\end{equation}
where $A = e, 5, \epsilon$. This means that the position of the peak
in $\omega$ for $\Im\sigma^\cB_5$ is the same as for
$\Im\sigma^\cB_e$.  As a consequence of that, the frequency dependence
of $\sigma_\e^\cB$ and $\sigma^\cB_5$ are remarkably close to each other
at high temperature, i.e. $T \gg \mu \,, \mu_5$, once they are
normalized to their respective zero frequency value. Eq.~(\ref{eq:ImsigmaT}) is valid modulo ${\cal O}\left(\frac{\mu^2}{T^2}\mu_5\right)$ and ${\cal O}\left(\frac{\mu_5^2}{T^2}\mu\right)$ corrections for $A = e$ and $A=5$ respectively, and exact for $A=\epsilon$.

\begin{center}
\begin{figure}[!]
\includegraphics[angle=0,height=0.3\textwidth]{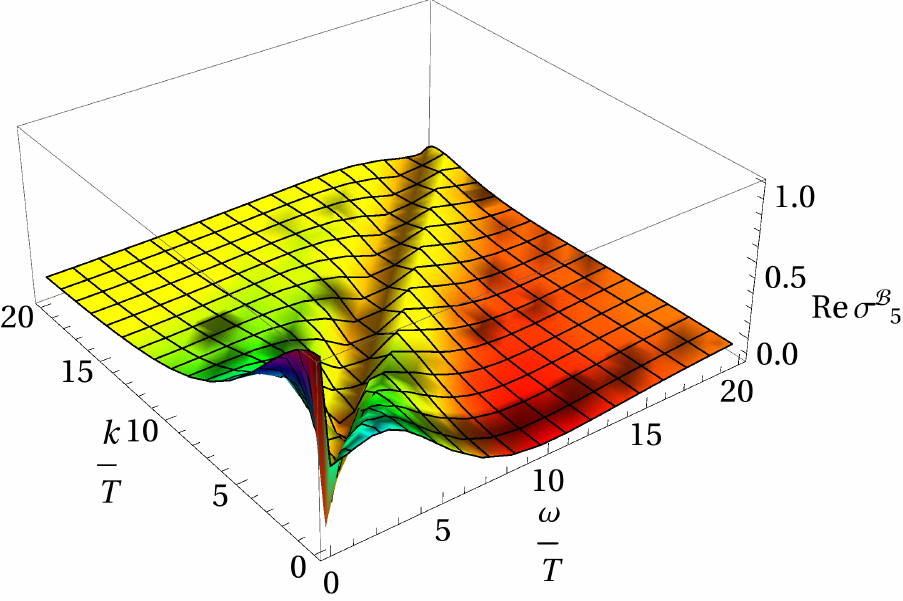}
\hspace{1cm}
\includegraphics[angle=0,height=0.3\textwidth]{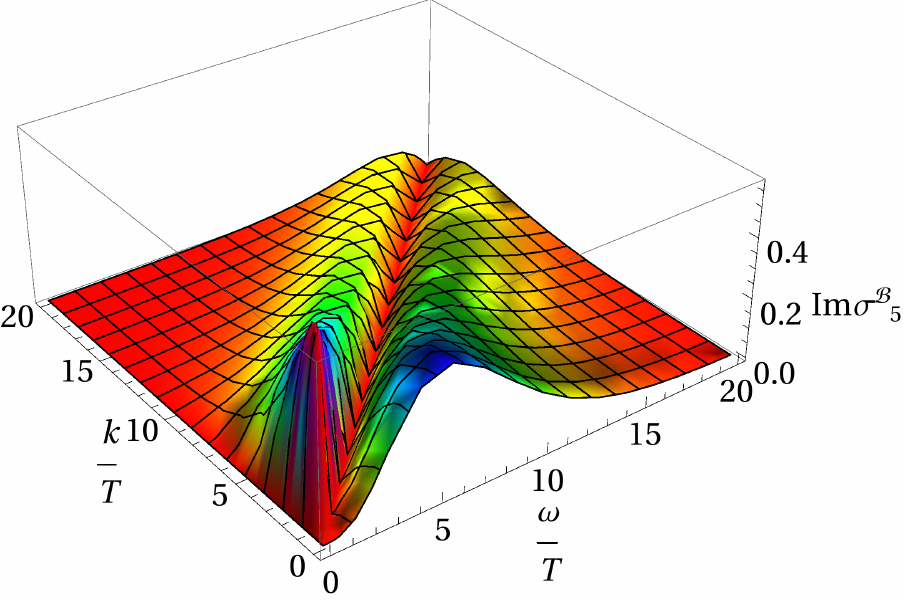}
\caption{Frequency and momentum dependence of the axial magnetic
  conductivity~$\sigma_5^\cB(\omega,k)$ normalized to its zero
  frequency-momentum value, from a numerical evaluation of
  Eqs.~(\ref{eq:GBA}) and (\ref{eq:fBA}). Left figure shows the real part,
  and right figure the imaginary part of the conductivity. We
  consider $\mu = 10 \,\textrm{MeV}$, $\mu_5 = 1 \,\textrm{MeV}$, and
temperature above the QCD phase transition $\tau = 440$. The plots of the vector
magnetic conductivity~$\sigma_v^\cB(\omega,k)$ are indistinguishable from these
ones at this temperature. [Color online]
}
\label{fig:3Dmagnetic5}
\end{figure}
\end{center}

We show in fig.~\ref{fig:3Dmagnetic5} the full frequency and momentum
dependence of $\sigma_5^\cB(\omega,k)$. Some of its features are
similar to the ones for vortical conductivities, see
Sec.~\ref{subsec:vortical-conductivity}, but there are some
differences. In particular: i) At high temperature, there is a peak at $k = 0$
and $\omega \simeq 5.406 T$, which tends to disappear when $k > 0$. ii) The
conductivities are not vanishing at $k=0$, $\omega \ne 0$, and they still
present a discontinuity at $k=0$, $\omega=0$.

The frequency and momentum dependence of all the other magnetic
conductivities are qualitatively similar to the ones described above,
so we do not show the corresponding plots. There are some extra
conductivities equivalent to the chiral magnetic ones, which are
associated with the presence of an external axial-magnetic field $\vec
B_5$. They follow from the correlators $\sigma^{\cB_5}_A \sim \langle
J_A J_5\rangle$, where $A = e, 5, \epsilon$. The study of these
conductivities is not of phenomenological interest in QCD, but they
might play a role in some condensed matter systems. It is
straightforward to check that in the free field theory of
Eqs.~(\ref{eq:Jv5}) and (\ref{eq:Delta_t}), the following relations apply
at one loop
\begin{equation}\label{eq:GB5s}
\widehat G^{\cB_5}_5(\omega,k) = \widehat G^\cB_e(\omega,k) \,, \qquad \widehat
G^{\cB_5}_e(\omega,k) = \widehat G^\cB_5(\omega,k) \,, \qquad \widehat
G^{\cB_5}_\epsilon(\omega,k) = \widehat G^\cV_5(\omega,k) \,.
\end{equation}
These identities just follow from the properties of the $\gamma$ matrices, in
particular $\gamma_5^2 = 1$, and the specific structure of the correlation
functions, c.f. Eq.~(\ref{eq:Gav}).

\subsection{Thermodynamic variables}

The pressure of the system can be computed from the correlator
$G^{0z,0z}(k) = \langle T^{0z}(k) T^{0z}(-k) \rangle$, which at one loop reads
\begin{eqnarray}
G^{0z,0z}(k_0,k) &=& -\frac{1}{16\beta}\sum_{\tilde\omega}
\int\frac{d^3q}{(2\pi)^3}
\tr \Bigg[ S(q) \gamma^0 S(q+k) \gamma^0 (2 q^z + k^z)^2 
+  2 S(q) \gamma^0 S(q+k) \gamma^z (2 i\tilde\omega + k^0 ) (2 q^z + k^z)
\nonumber \\
&&+ S(q) \gamma^z S(q+k) \gamma^z (2 i\tilde\omega + k^0 )^2 \Bigg] 
+ \frac{3}{4} \left( \langle T^{zz}\rangle -  \langle T^{00} \rangle \right) \,.
\label{eq:Gz}
\end{eqnarray}
The last term is the contribution coming from the seagull diagram,
see~\cite{Manes:2012hf}. The precise relation with the pressure reads
\begin{equation}
P=\lim_{\vec{k}\to 0} G^{0z,0z}(k_0,k) \big|_{k_0 = 0}  \,.
\end{equation}
After an evaluation of Eq.~(\ref{eq:Gz}), and considering the limits at zero
frequency and momentum, the result for the pressure in the free theory reads
\begin{equation}
P = \frac{1}{6\pi^2} \int_0^\infty dq \, q^3 \, f_P(q) =
\frac{1}{12\pi^2} \left[ \mu^4+ 6\mu^2\mu_5^2 + \mu_5^4 + 2\pi^2 T^2 (\mu^2 +
\mu_5^2) +  \frac{7}{15}\pi^4 T^4 \right] \,, \label{eq:T0zT0z} 
\end{equation}
where $f_P(q) =   \sum_{s,t=\pm} n(E_q+t\mu_s)$. This result corresponds to the
pressure of an ideal gas of
massless fermions of spin $1/2$ at finite temperature and chiral chemical
potentials. By considering $\mu_5=0$ in the previous formula, we recover the
standard result in the literature~\cite{LeBellac:1996}. From this expression one
may obtain the rest of thermodynamical quantities, in particular the energy
density $\epsilon = 3P$, entropy density
\begin{equation}
s = \frac{\partial P}{\partial T} = \frac{T}{45} \left( 15(\mu^2 + \mu_5^2) +
7\pi^2 T^2 \right) \,,
\end{equation}
and the baryon and axial densities, respectively,
\begin{eqnarray}
\rho_e = \frac{\partial P}{\partial \mu} &=& \frac{\mu}{3\pi^2}\left( \mu^2 +
3\mu_5^2 + \pi^2 T^2\right) \,, \\
\rho_5 = \frac{\partial P}{\partial \mu_5} &=&  \frac{\mu_5}{3\pi^2}\left( 3\mu^2 +
\mu_5^2 + \pi^2 T^2\right) \,.
\end{eqnarray}
It is easy to check that previous relations fulfill
\begin{equation}
\epsilon + P = T s + \mu n_e + \mu_5 n_5 \,.
\end{equation}
The pressure can be obtained also from a direct computation of the
thermodynamical potential of a free gas of fermions with chiral chemical
potential, as it was done in~\cite{Fukushima:2008xe}. Finally, notice that
Eq.~(\ref{eq:T0zT0z}) can be expressed as a sum
of right-handed and left-handed fermionic species contributing to the
pressure, $P = P_+ + P_-$, where
\begin{equation}
P_{\pm} = \frac{1}{24\pi^2} \left[ \mu_\pm^4 + 2\pi^2 T^2 \mu_\pm^2  + 
\frac{7}{15}\pi^4 T^4 \right] \,.
\end{equation}

\section{Strongly coupled regime}
\label{sec:strongly_coupled_regime}

To study the frequency dependence of a strongly coupled plasma we will
use a holographic model similar to the one introduced in
\cite{Landsteiner:2011iq}. This model implements the  gauge and mixed
gauge-gravitational triangle anomalies. The difference of the present case with
the model of \cite{Landsteiner:2011iq} is the inclusion of a conserved current
which will be interpreted as the electric current beside the non conserved axial
current. The presence of both currents in the model allows us to compute the
frequency dependence of the chiral magnetic~\cite{Yee:2009vw}, separation and
vortical effects.

The dual model consists on a five dimensional gravity theory with two gauge
fields $U(1)_V\times U(1)_A$. The action writes 
\begin{eqnarray}
\nonumber S &=& \frac{1}{16\pi G} \int d^5x \sqrt{-g} \left[ R +   12 - \frac 1
4 \left( F_{MN} F^{MN} +  F^{(5)}_{MN} F^{(5)MN}\right) + \right. \\
&&\left.+ \epsilon^{MNPQR} 
A^{(5)}_{M} \left(\frac\kappa 3 F^{(5)}_{NP} F^{(5)}_{QR}+ \kappa  F_{NP} F_{QR}
+ \lambda R^A\,_{BNP}   R^B\,_{AQR} \right) \right] + S_{GH} + S_{CSK}\,, \label{eq:S}\\ 
   S_{GH} &=& \frac{1}{8\pi G}\int_{\partial} d^4x \sqrt{-h} \, K \,,
\label{eq:S_GH}\\ 
  S_{CSK} &=& - \frac{1}{2\pi G} \int_{\partial} d^4x \sqrt{-h} \, \lambda n_M
\epsilon^{MNPQR} A^{(5)}_N K_{PL} D_Q K_R^L \,, \label{eq:S_CSK} 
\end{eqnarray}
where $n_A$ is a normal vector to the $AdS$ boundary and $K_{MN}$ is the
extrinsic curvature. In addition to this action it is necessary to include a
boundary counterterm in order to make it
finite.~\footnote{See~\cite{Landsteiner:2011iq} for a 
detailed discussion on the holographic renormalization of the model,
and~\cite{Landsteiner:2012dm}  for the need of inclusion of $S_{CSK}$.}
This action is invariant under diffeomorphisms and vector gauge transformations,
but it is not invariant under axial gauge transformations. The variation of the
action under the latter shows results in the axial anomaly
\begin{equation}
\delta_{\xi_5} ( S + S_{GH} + S_{CSK} ) = \frac{1}{16\pi G} \int_\partial d^4 x
\sqrt{-g}\,\xi_5 \epsilon^{\mu \nu \rho \beta}\left( \frac{\kappa}{3}
F^{(5)}_{\mu \nu} F^{(5)}_{\rho\beta} + \kappa F_{\mu \nu} F_{\rho\beta} +
\lambda  R^{\alpha}\,_{\delta\mu\nu} R^{\delta}\,_{\alpha\rho\beta}\right)\,.
\end{equation}
This expression allows us to fix the value of the $\kappa$ and $\lambda$
parameters in terms of the anomalous coefficients of the field theory, so that
\begin{equation}
\label{eq:kappa-lambda_value}
\frac{1}{16\pi G}\kappa = -\frac{1}{16\pi^2} \qquad , \qquad \frac{1}{16\pi G}
\lambda = -\frac{1}{384\pi^2}.
\end{equation} 
This system admits a static charged black hole solution
\begin{eqnarray}
ds^2 &=& r^2\left(-f(r)dt^2+d\vec x^2\right) +\frac{dr^2}{r^2f(r)}\quad\,,\quad
A = - \frac{\mu }{r^2}dt \quad\, , \quad A^{(5)} = - \frac{\mu_5 }{r^2}dt \, ,
\end{eqnarray}
where  $f(r)=1-\frac{m }{r^4} + \frac{q^2 }{r^6 }$ is the blackening factor,
while the mass and charge of the black hole are defined, respectively,
as~\footnote{Notice that we have set the $AdS$ and
black hole horizon radius to one.}
\begin{eqnarray}
m &=& 1+ q^2 \quad \,, \quad q^2 = \frac{\mu^2 +\mu_5^2 }{3}\,.
\end{eqnarray}
The Hawking temperature is given by $T=\frac{2m-3q^2}{2\pi }$. The
extremal solution is obtained when $q=\sqrt{2}$. With these ingredients one can
compute the pressure of the holographic model, and it reads~\footnote{See
e.g.~\cite{Erdmenger:2008rm,Megias:2013joa} for the result with $\mu=0$ and
$\mu_5 \ne 0$.}
\begin{equation}
P = \frac{m}{16\pi G} = \frac{1}{16\pi G} \left(\frac{\pi T}{2}\right)^4 \left(
\sqrt{1 + \frac{2}{3}\frac{(\mu^2 + \mu_5^2)}{\pi^2 T^2}} + 1 \right)^3 \left(3
\sqrt{1 + \frac{2}{3}\frac{(\mu^2 + \mu_5^2)}{\pi^2 T^2}} -1  \right) \,.
\end{equation}
As expected this result is different from the one obtained in the free gas of
chiral fermions, cf. Eq.~(\ref{eq:T0zT0z}).

Our purpose is to compute two point retarded correlators in a linear response
regime on top of this equilibrium background. To do so we need to introduce
fluctuations of the gauge fields and metric components
$A(t,y,r)=A_{(0)}(r)+\epsilon \, a(t,y,r)$,
$A^{(5)}(t,y,r)=A_{(0)}^{(5)}(r) + \epsilon \, \tilde a(t,y,r)$ and
$g_{MN}(t,y,r)=g^{(0)}_{MN}(r)+\epsilon  \, h_{MN}(t,y,r)$. The nature of the
correlators we want to compute tells us that it is enough to study only the
shear sector. Allowing a $y$ dependence induces a breaking of rotational
symmetry to the $SO(2)$ group around the axis defined by the $y$ coordinate.
Therefore to study the shear sector it is enough to switch on the components  
\begin{eqnarray}
\Phi^T = (a_x, \tilde a_x, h_{t}^x, h_{y}^x, a_z, \tilde a_z, h_{t}^z,
h_{y}^z)^T \,.
\end{eqnarray}
We will consider the gauge fixing $A_r=A^{(5)}_r=h_{rM}=0$. After
introducing in the action this ansatz and taking
variations with respect to the fluctuations we get the linearized
equation of motions for the system,
\begin{eqnarray}
\label{eq_As} 0&=& a_i'' ( u )+\frac { f'  }{f}a_i'( u )+ \frac {1  }{4uf^2
}\left(\omega^2-  f
{k}^{2} \right)a_i( u) -  \frac{ \mu }{f}h'^{i}_t ( u ) + \frac{  4i k}{f} 
\kappa\epsilon_{ij}\left(\mu  \tilde a_j(u) + \mu_5 a_j(u)\right) \,,\\
\nonumber 0&=& \tilde a_i'' ( u )+\frac { f' }{f}\tilde a_i'( u )+ \frac {1 
}{4uf^2 }\left(\omega^2-  f
{k}^{2} \right)\tilde a_i( u) -  \frac{\mu_5 }{f}h'^{i}_t ( u ) + \frac{  4i
k}{f}  \epsilon_{ij}\left(\kappa(\mu  a_j(u) + \mu_5 a_j(u)) +\lambda\frac{3  
\left(u^3\mu_T ^2 + 2(f-1)\right) }{u}h'^j_t(u)\right) \,,\\
\label{eq_tAs}\\
\label{eq_Hts}\nonumber 0&=&  h''^{i}_t(u) - \frac{h'^{i}_t(u)}{u} -\frac {1}{4
uf}\left(k^2 h^i_t(u)+\omega k h^i_y \left( u \right)  \right) - u(\mu_5  \tilde
a_i(u) + \mu   a_i(u))'
+ i\lambda u k \epsilon_{ij}\left[\frac{2  k^2  \mu_5 }{f}h_t^j(u)  -8   \mu_5(u
h'^{j}_t(u))' \right.\\
&&\left.+ \frac{2  k  \omega \mu_5 }{f}h_y^j(u) -\frac{8}{3}    \left((7 u-3)
\mu_T ^2+8 u \mu_T^2-9\right) \tilde a_j(u)-4  u  \left( (5u-2) \mu_T
^2-6\right) \tilde a'_{j}(u)
\right] \,,\\
\label{eq_Hys}\nonumber 0&=& h''^{i}_y(u)+
\frac{\left(f/u\right)'}{f/u}h'^{i}_y(u)+\frac{1}{4u
f^2}\left(\omega^2h_y^i(u)+\omega k h_t^i(u)\right)+
2uik\lambda\mu_5\epsilon_{ij}\left[ -\frac{ 1}{f^2}( k  \omega   h_t^j(u)+
\omega^2 h_z^j(u)) -4  u h''^j_y(u)\right.\\
 && \left.-4\left(3  +\frac{ u^3 \mu_T ^2 -6}{3 f}\right) h'^j_y(u) \right] \,,
\end{eqnarray}
and the constraints
\begin{eqnarray}
\label{constraints} 0&=&\omega h'^i_t + k f h'^i_y   -u \omega( \mu_5  \tilde
a_i+ \mu  a_i) - 4i k \lambda\epsilon_{ij}  \left(\omega \left(6(f-1)  +3  u^3 
\mu_T ^2\right)\tilde a_j + 2 \mu_5 u^2( \omega  h'^j_t+  k   f h'^j_y)\right)
\,,
\end{eqnarray}
where we have redefined $\mu_T^2\equiv \mu^2+\mu_5^2$, $u=1/r^2$ and $i=x,z$.
Notice that we also have Fourier transformed the 
fields.

As we are interested in computing retarded propagators we need to impose
infalling boundary conditions at the horizon. This is the main boundary
condition that must be satisfied. Since the system is second order we
need a second boundary condition to specify a unique solution. This is done by
demanding 
that the matrix of linearly independent solutions
goes to the unit matrix at the boundary. These boundary conditions define the
bulk-to-boundary propagator.
In \cite{Son:2002sd,Kaminski:2009dh} a prescription to
obtain two point functions in holography is discussed. The procedure is as
follows: first we
have to expand the renormalized action up to second order in the perturbative
parameter $\epsilon$, and then Fourier transform it to get an expression of the
form 
\be \label{eq:2ndor}
\delta S^{(2)}=\int \frac{\dd^d k}{(2\pi)^d} \lbrace \Phi^I_{-k} \cA_{IJ} \Phi
'^J_k + \Phi^I_{-k}  \cB_{IJ} \Phi^J_k \rbrace\Big{|}_{r\to\infty}\,.
\ee
On the other hand, we have to find a maximal set of linearly independent
solutions satisfying ingoing boundary conditions and build the matrix
$H^I_J(k,u)$ where
each column consists of a solution of the linearly independent set. Finally the
desired solution with the boundary sources switched on is
\be\label{eq:f}
\Phi^I_k (u) = F^I\,_J (k,u)\, \varphi^J_k\,,
\ee
where $F=H(k,u)H^{-1}(k,0)$ and $\varphi^J_k$ are the sources of the dual field
theory. From this we can read the retarded correlators which look like
\be\label{eq:GR}
G_{IJ}(k)= -2 \lim_{u\to 0} \left(\cA_{IM} (F^M\,_J (k,u))' +\cB_{IJ}\right)\,.
\ee
After some tedious computation we can extract the matrices $\cA$ and $\cB$ for
our system
\be
\cA=\frac{1}{16\pi G}\mathrm{Diag}\left(
f(u), f(u), -\frac{1}{u},\frac{f(u)}{u}, f(u), f(u) ,
-\frac{1}{u},\frac{f(u)}{u} \,
\right)
\ee
and
\be 
\cB=
\frac{1}{16\pi G}\left(
\begin{array}{cc}
 B & 0_{4\times 4}   \\
 0_{4\times 4} &  B  
\end{array}
\right)\quad,\quad B=
-\frac{1}{2}\left(
\begin{array}{cccc}
 \left(k^2-\omega^2\right) \log u & 0 & 0 & 0  \\
 0 &  \left(k^2-\omega^2\right) \log u & -4\mu_5 & 0  \\
 0 & 0 & \frac{ k^2}{u}+\frac{12}{u^2}& \frac{ k \omega}{u }  \\
 0 & 0 & \frac{ k \omega}{u} & \frac{ \omega^2}{u}-\frac{12 }{u^2} 
\end{array}
\right) \,.
\ee

Now we have all the ingredients to compute the retarded Green
functions and to use the Kubo formulae
(\ref{eq:Kubo1}) and (\ref{eq:Kubo2}) to extract the anomaly induced
transport coefficients. The zero frequency case can be done
analytically by setting $\omega=0$ and looking for a linearized
solution in the momentum $k$ (to see a detailed way of solving the
system see \cite{Yee:2009vw,Gynther:2010ed,Amado:2011zx,Landsteiner:2011iq}). The
conductivities in this case are
\begin{eqnarray}
 \sigma^\cB_{A,(0)} =  \frac{ 1}{2 \pi ^2}\left\{
 \begin{array}{l}
   \mu_5\\
   \mu\\
   \mu\mu_5
   \end{array}
   \right.
  \quad ,\quad \sigma^{\cB_5}_{A,(0)} = \frac{1 }{2 \pi ^2}\left\{
  \begin{array}{l}
    \mu \\
    \mu_5 \\
    \frac{\mu^2+\mu_5^2}{2}+\frac{\pi^2T^2}{6}
  \end{array}
  \right.
  \quad  ,\quad \sigma^\cV_{A,(0)} = \frac{ 1}{2 \pi ^2}\left\{
  \begin{array}{l}
    \mu\mu_5\\
     \frac{\mu^2+\mu_5^2}{2}+\frac{\pi^2T^2}{6}\\
     \mu_5\left( \frac{\mu^2+\mu_5^2}{3}+\frac{\pi^2T^2}{3}\right)
  \end{array}
  \right.\,.
  \end{eqnarray}
To study the frequency dependence we have to resort to numerics. The system of
differential equations presents a singularity at $u=1$, so we have to implement
a methodology to integrate the equations from this point to the boundary. As a
first step we redefine the fields to ensure the infalling boundary condition
\bea
a^i(u) &=& (1-u)^{-i w}\,b^i(u)\,,\\
\tilde a^i(u) &=& (1-u)^{-i w}\,c^i(u)\,,\\
h^i_t(u) &=& (1-u)^{-i w+1}\,H^i_t(u)\,,\\
h^i_y(u) &=& (1-u)^{-i w}\,H^i_y(u)\,,
\eea
where $(w,P)=(\omega,k)/4\pi T$. Now the infalling condition is translated to a
regularity condition on the fields $(b^i,c^i,H^i_t,H^i_y)$. Then we have to find
eight linearly independent solutions to construct the matrix $F$, but the system
is subject to two constraints reminding us that not all the fields are 
independent. Substituting these redefined fields into the constraints and
evaluating them at the horizon, it is possible to find the  relation
\begin{eqnarray}
\nonumber H_y^i(1) &=&  \frac{3  (i+w )}{P \left(\mu_T ^2-6\right)}H_t^i(1) -
\frac{27 i  \mu b^i(1) + i  \mu_5 \left(3+128 P^2 \lambda ^2 \left(\mu_T
^2-6\right)^2 \left(\mu_T ^2-2\right)\right)c^i(1)}{P \left(\mu_T ^2-6\right)
\left(256 P^2 \lambda ^2 \mu_5^2 \left(\mu_T ^2-6\right)^2-9\right)} \\
&& -\lambda\epsilon_{ij}\frac{16    \mu  \mu_5b^j(1) + 72    \left(\mu_T
^2-6\right)^2c^j(1)}{256 P^2 \lambda ^2 \mu_5^2 \left(\mu_T ^2-6\right)^2-9}.
\end{eqnarray} 
This formula makes clear that we only have freedom to fix the six
values $(H_t^i(1),b^i(1),c^i(1))$, while the remaining two are given by pure
gauge solutions arising from gauge transformations of 
the trivial solution. We choose them to be
\bea
\nonumber \Phi(1)=\begin{array}{cccccccc}
\left(\begin{array}{c}
1\\
0\\
0\\
H_y^x(1)\\
0\\
0\\
0\\
H_y^z(1)
\end{array}\right) ,&
\left(\begin{array}{c}
 0\\
 1\\
 0\\
 H_y^x(1)\\
 0\\
 0\\
 0\\
 H_y^z(1)
 \end{array}\right), &
\left(\begin{array}{c}
 0\\
 0\\
 1\\
 H_y^x(1)\\
 0\\
 0\\
 0\\
 H_y^z(1)
 \end{array}\right), &
 \left(\begin{array}{c}
0\\
0\\
w\\
-p\\
0\\
0\\
0\\
0
\end{array}\right) \,,&
\left(\begin{array}{c}
 0\\
 0\\
 0\\
  H_y^x(1)\\
 1\\
 0\\
 0\\
 H_y^z(1)
 \end{array}\right), &
\left(\begin{array}{c}
 0\\
 0\\
 0\\
  H_y^x(1)\\
 0\\
 1\\
 0\\
 H_y^z(1)
 \end{array}\right),&
\left(\begin{array}{c}
 0\\
 0\\
 0\\
  H_y^x(1)\\
 0\\
 0\\
 1\\
 H_y^z(1)
 \end{array}\right),&
\left(\begin{array}{c}
0\\
0\\ 
 0\\
 0\\
 0\\
 0\\
 w\\
 -p
 \end{array}\right)\,.
 \end{array}\\
 \eea
 
 With these first eight vectors we can find numerically the linearly
 independent solutions. The matrix $H$ is built up as
 $H^I_J(u)=(\Phi^I(u))_J $. For the numerical computation we use the
 values $\kappa = 1$ and $\lambda = 1/24$ since we know from
 Eq. (\ref{eq:kappa-lambda_value}) the ratio $\lambda /\kappa =
 1/24$.  
 
 As the Kubo formulae demand to take the limit $k\to 0$ we
 fixed the infrared cutoff momentum $P_c=1/1000$.  We study first the
 case of interest for heavy ion collisions, that corresponds to the
 high temperature situation.  In particular if we assume the vector
 chemical potential is of order $10\,$MeV and a temperature around
 the QCD critical temperature $T_c=160\,$MeV, that fixes
 $\tau=95$. For temperatures of order $700\,$MeV we get $\tau=440$. In
 what follows we will analyze the theory for these two particular
 temperatures.
 
\begin{center}
\begin{figure}[t]
 \includegraphics[angle=0,width=0.4\textwidth]{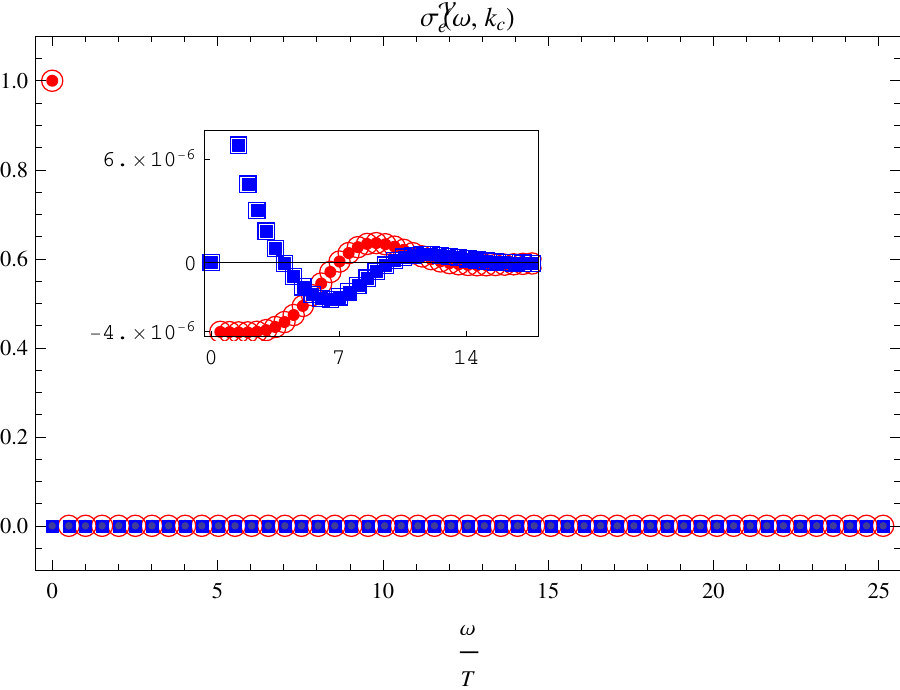} \hspace{1cm}
 \includegraphics[angle=0,width=0.4\textwidth]{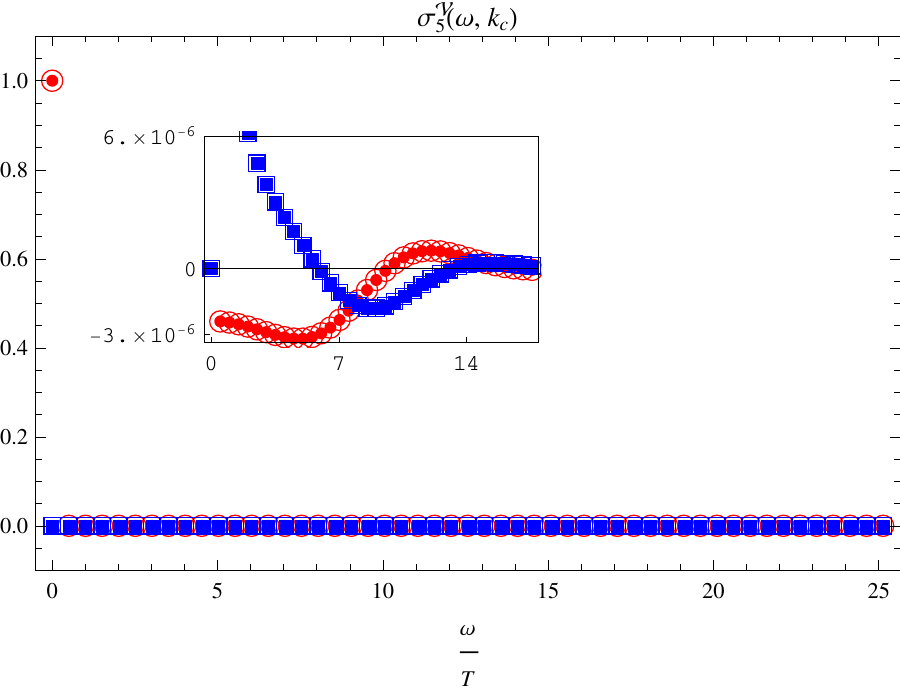} \\
 \includegraphics[angle=0,width=0.4\textwidth]{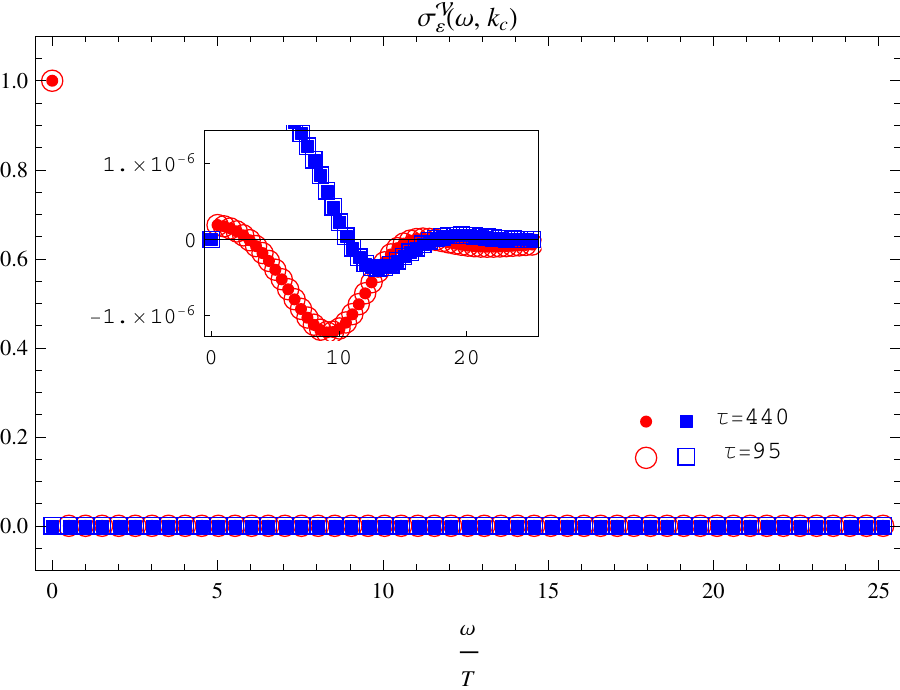}
\hspace{1cm} 
 \includegraphics[angle=0,width=0.4\textwidth]{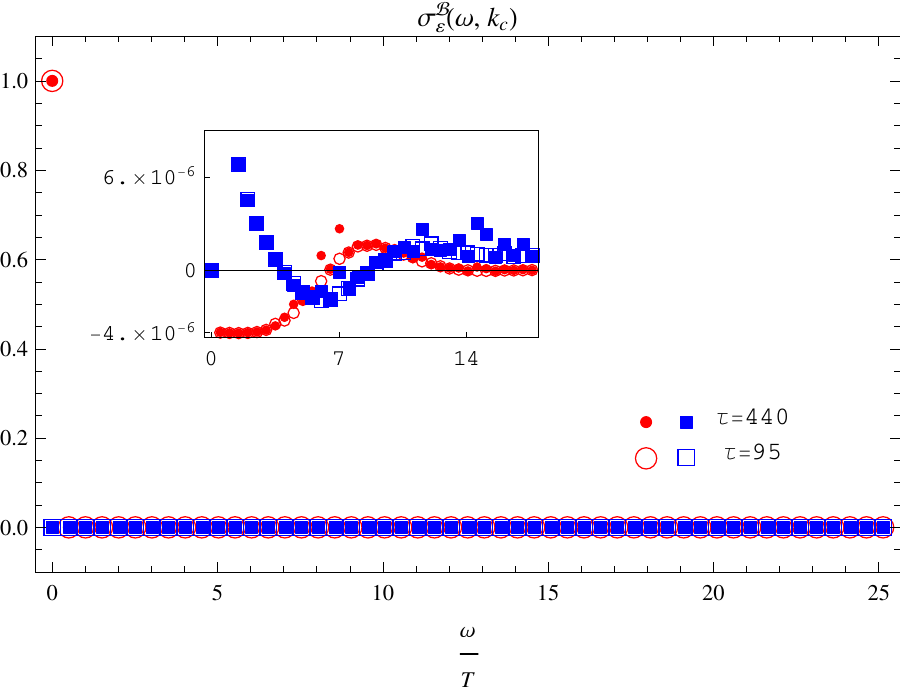}
\caption{Frequency dependence of the chiral vortical conductivities
  and chiral magnetic energy conductivity for finite chemical
  potentials and temperature around the QCD critical value ($\tau =
  95$, $\mu_5/\mu=0.008$) and temperature above the phase transition
  ($\tau = 440$, $\mu_5/\mu=0.03$). Red circles represent the real part
  of the conductivities and blue squares the imaginary part. [Color
    online]}
\label{fig:sigmaVStrong}
\end{figure}
\end{center}

\begin{center}
\begin{figure}[t]
\includegraphics[angle=0,width=0.4\textwidth]{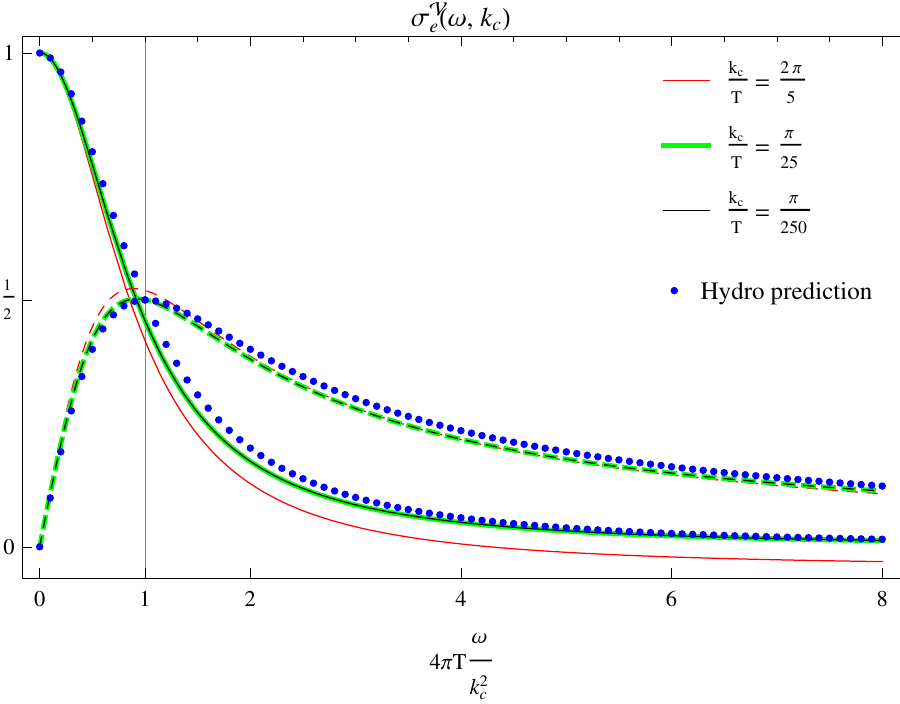}
\includegraphics[angle=0,width=0.4\textwidth]{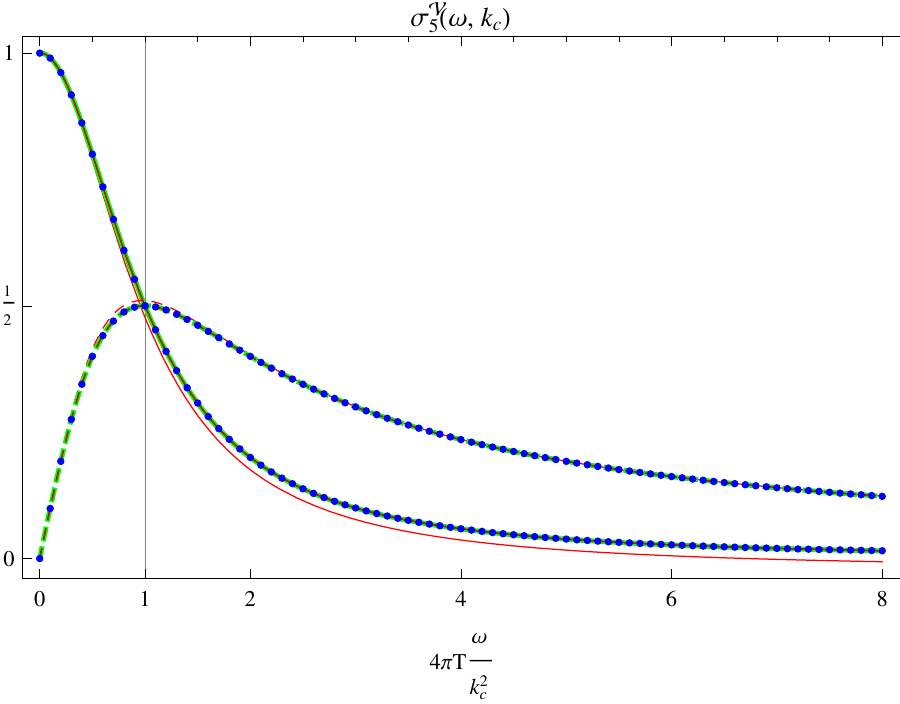}\\
\includegraphics[angle=0,width=0.39\textwidth]{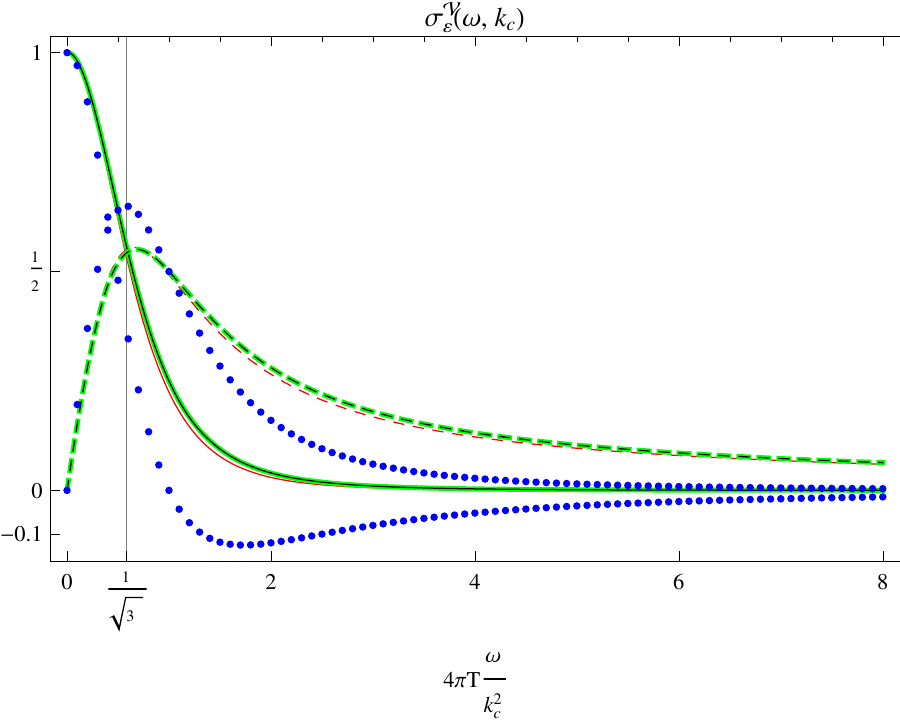}
\includegraphics[angle=0,width=0.4\textwidth]{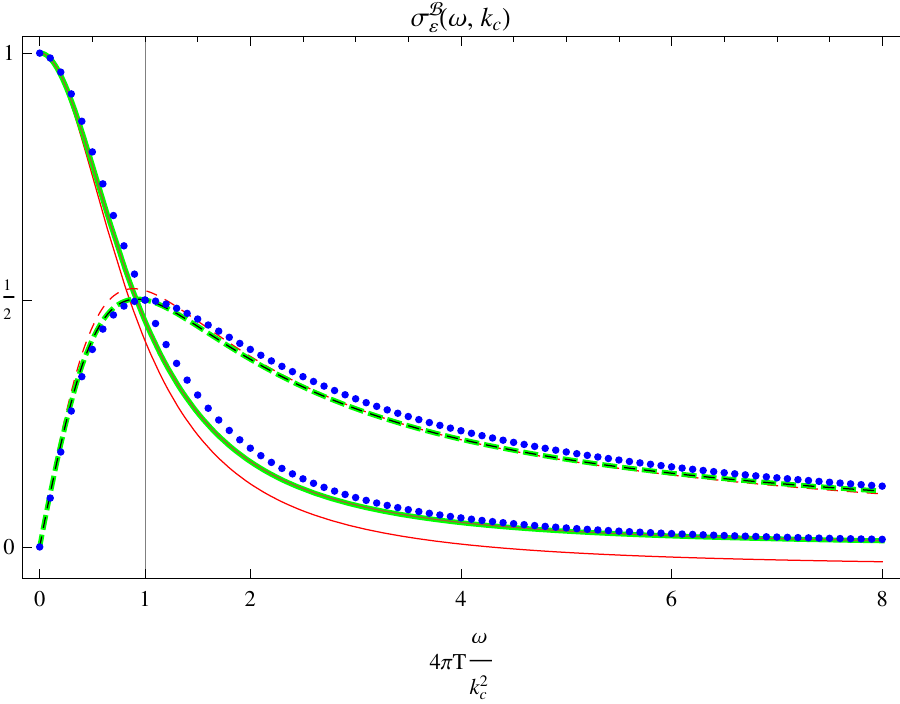}
\caption{Study of the width of the vortical conductivities with the
  infrared momentum cut-off used for the numerics.  Solid and dashed
  lines show the real and imaginary parts, respectively. Red lines show
  $P_c=1/10$, green $P_c=1/100$ and black $P_c=1/1000$.  The parameters are
  ($\tau = 95$, $\mu_5/\mu=0.008$). Dots are the hydrodynamical fit (see section
\ref{sec:hydro}) and the vertical lines represent the positions of the maxima
predicted by hydrodynamics. Notice that the regime in frequencies in
this figure is much smaller than the one in figure~\ref{fig:sigmaVStrong}.
[Color online]}
\label{fig:sigmaVcompare1}
\end{figure}
\end{center}

 Figures \ref{fig:sigmaVStrong} and \ref{fig:sigmaVcompare1} show
the behavior of the chiral vortical
conductivities. Figure~\ref{fig:sigmaVStrong} is quite similar to the
weakly coupled behavior. Once the frequency moves away from zero, the
conductivity in the strongly coupled regime drops 6 orders of
magnitude and shows a damped oscillation. In the free fermion case we
have seen that the vortical conductivities are defined as piecewise
functions of the frequency, when the source is a homogeneous function
of the space coordinates. However in the present case we are obtaining
fast decaying functions of the frequency but with a non vanishing
width.  Therefore to study whether
the strongly coupled coefficients have the decay smoothed by the
strong interaction, or whether we see a width as a consequence of the
small but non vanishing momentum used for the numerics, we compute the
conductivities evaluating them at three infrared cut-off momenta for
temperature $\tau = 95$ and chemical potentials
$\mu_5/\mu=0.008$. Then in figure \ref{fig:sigmaVcompare1} we plot all
the fast decaying conductivities as a function of $4\pi T\omega/k_c^2$
for the dimensionless momentum values $P_c=1/10,1/100,1/1000$. Analyzing
the plot we realize that the approximate position of either the peaks
in the imaginary part or the width in the real part is of order
$\omega\sim k_c^2/(4\pi T)$. Then we can infer that the conductivities
for a homogeneous source vanish at $\omega\neq 0$, and they are
discontinuous functions at $\omega = 0$, coinciding exactly with the
weakly coupled conductivities. Notice that the constant $1/(4\pi T)$ is the
shear diffusion constant, this number suggests that this small frequency and
momentum behavior are governed by hydrodynamics This point will be addressed
in section \ref{sec:hydro}.

\begin{center}
\begin{figure}[t!]
\includegraphics[angle=0,width=0.65\textwidth]{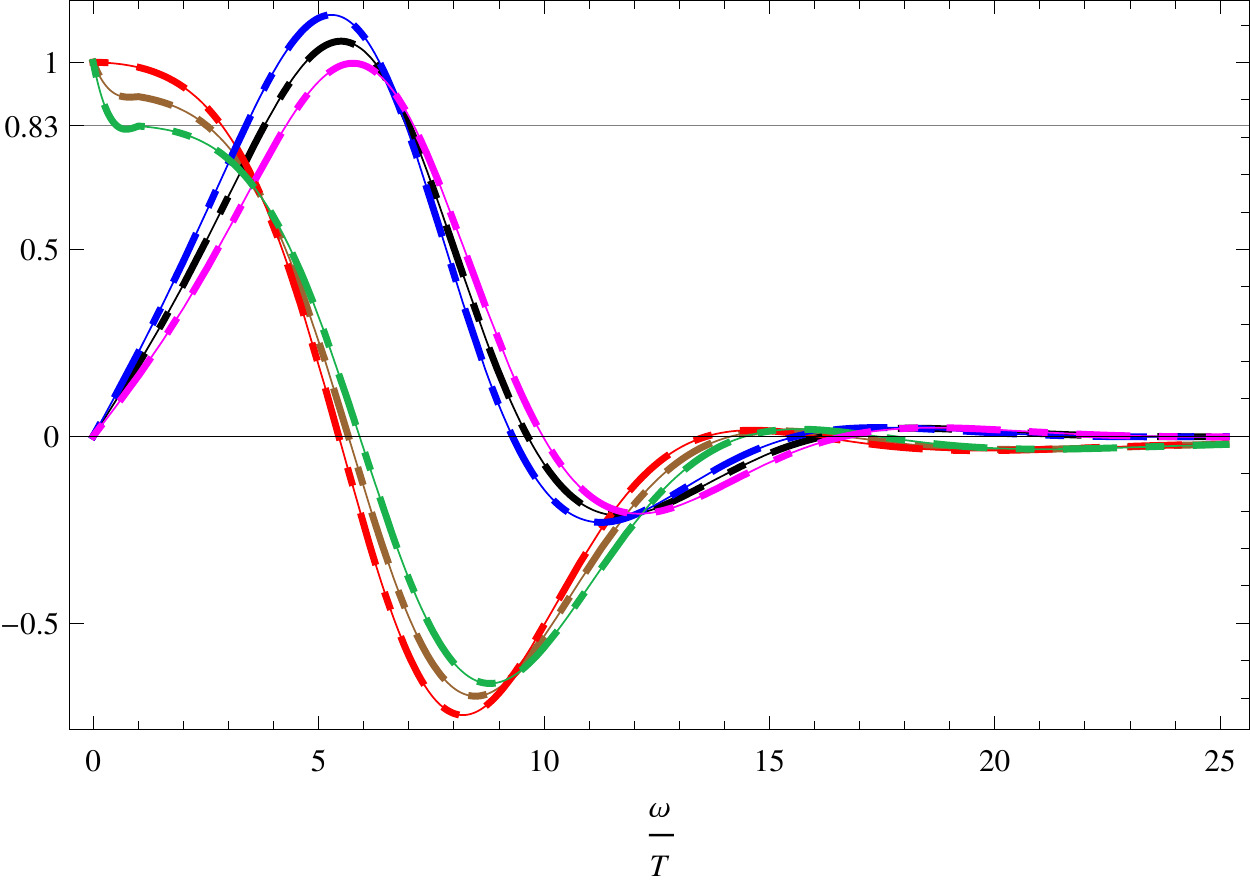}
\includegraphics[angle=0,width=0.34\textwidth]{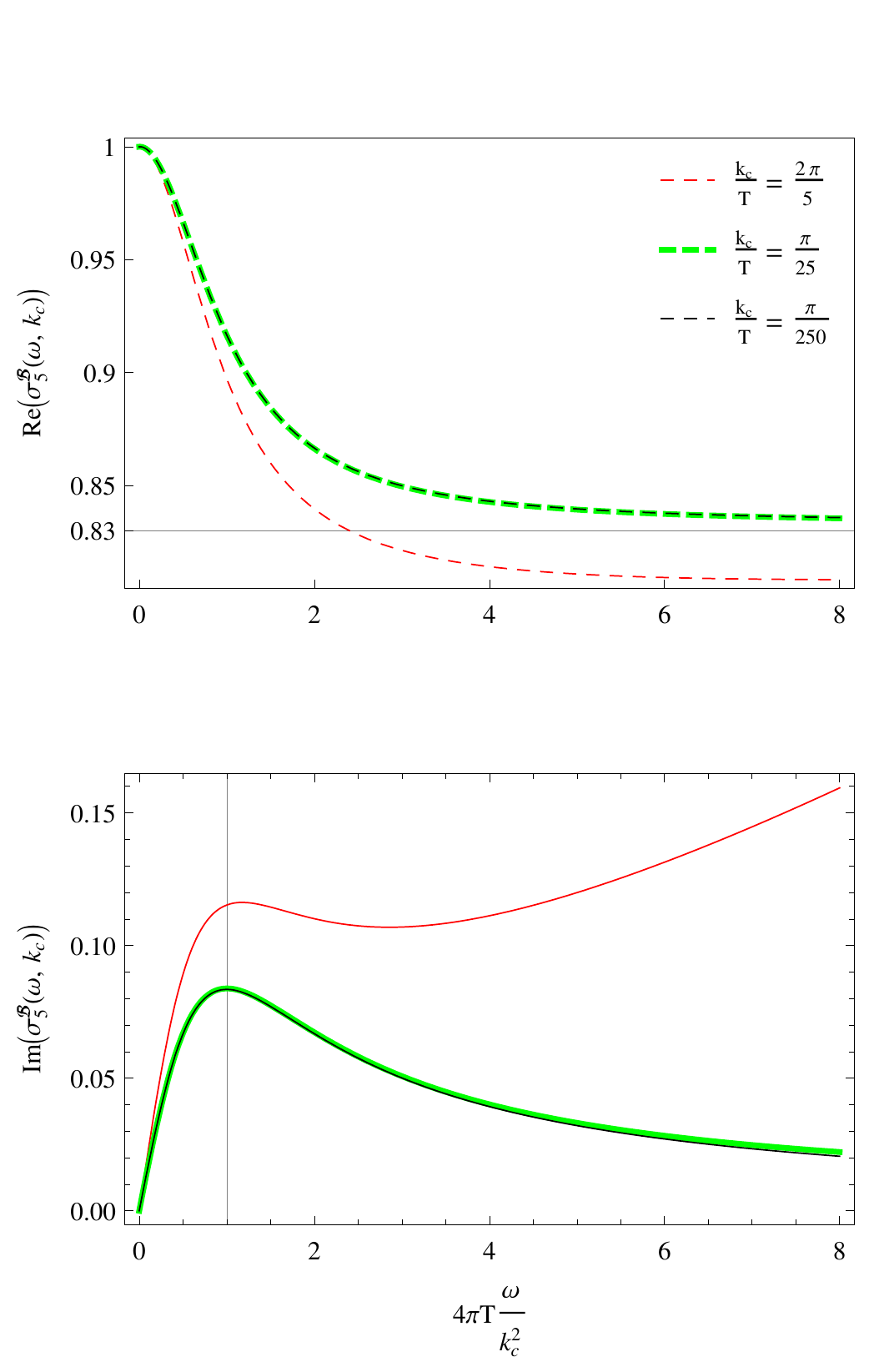}
\caption{Frequency dependence of the chiral magnetic conductivities
for high temperatures. Thick dashed lines correspond to $\tau=440$ and continuous
lines to $\tau=95$. 
Red and blue are real and imaginary parts of $\sigma_e^\cB$ respectively, brown
and black lines real and 
imaginary parts of $\sigma_5^\cB$, green and magenta lines are real and imaginary part
of $\sigma_5^{\cB_5}$. [Color online]}
\label{fig:sigmaBStrong}
\end{figure}
\end{center}

\begin{center}
\begin{figure}[t!]
\includegraphics[angle=0,width=0.495\textwidth]{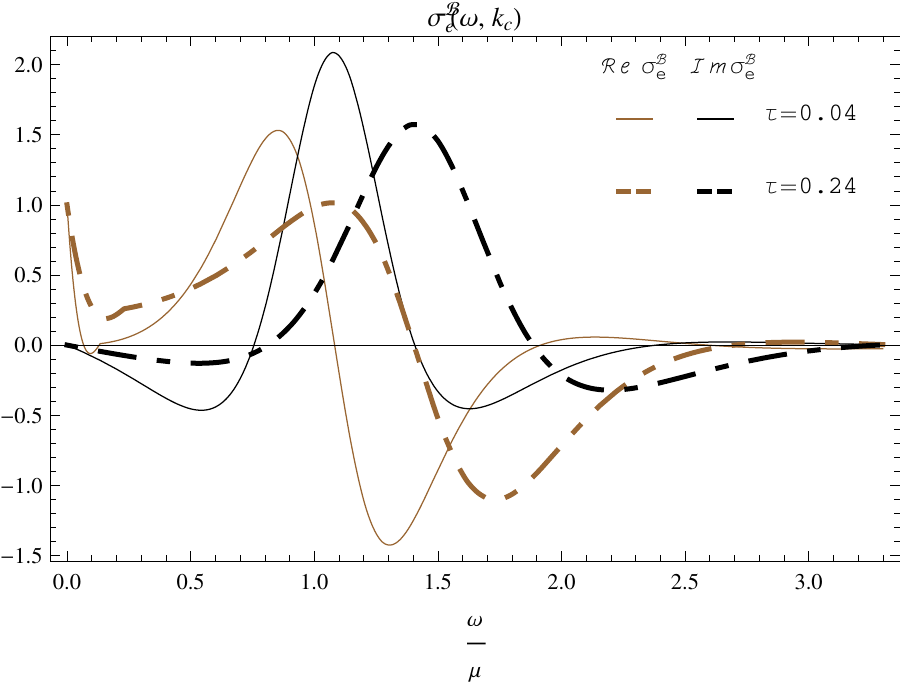} 
\includegraphics[angle=0,width=0.495\textwidth]{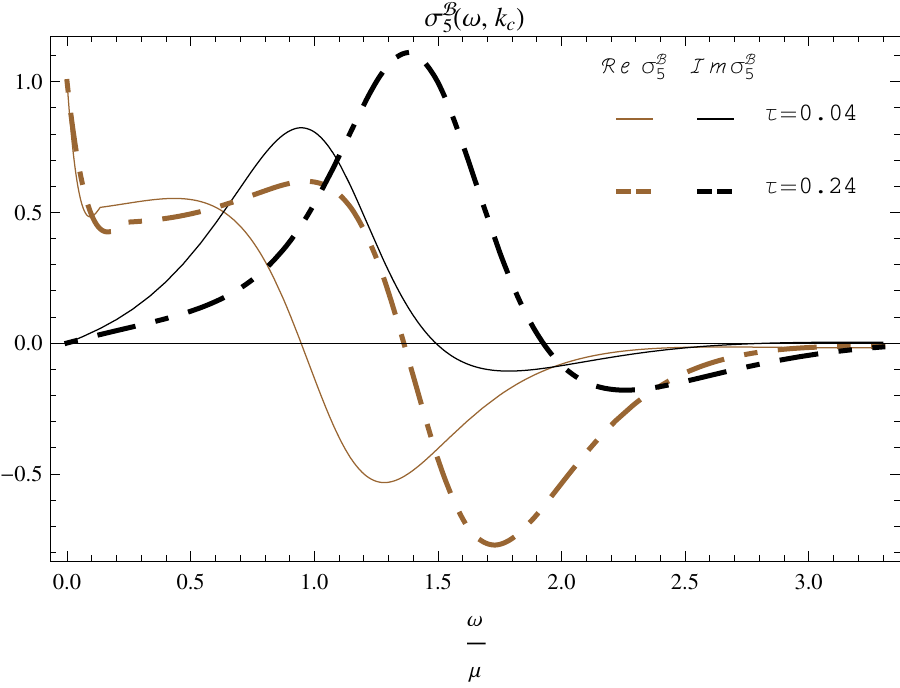} 
\caption{Frequency dependence of the chiral magnetic conductivities
for temperatures close to zero, ($\tau = 0.04$, $\mu_5/\mu=7\times 10^{-5}$) and
($\tau = 0.24$, $\mu_5/\mu=0.7$).}
\label{fig:sigmaBStronglow}
\end{figure}
\end{center}

\begin{center}
\begin{figure}[t!]
\includegraphics[angle=0,height=0.3\textwidth]{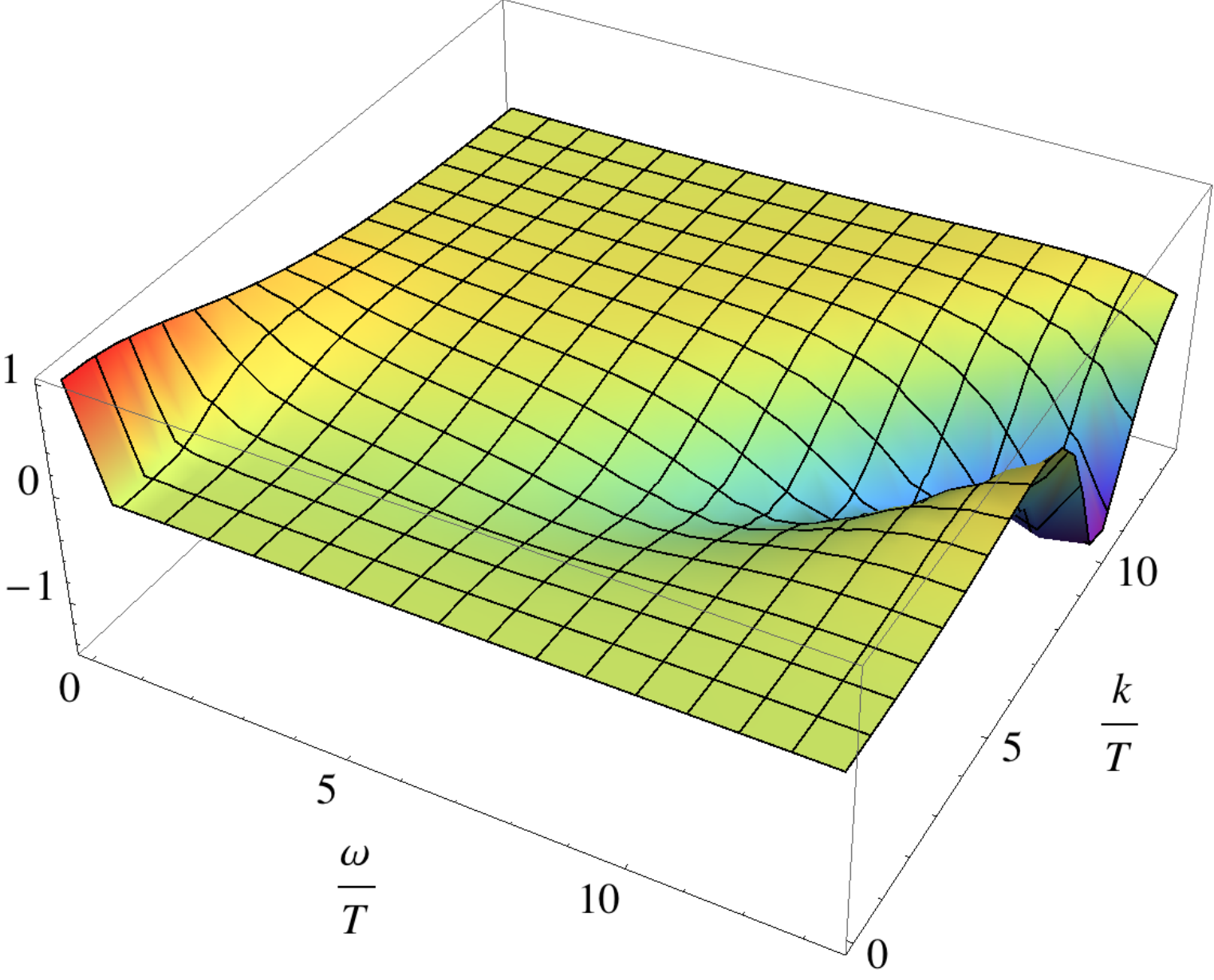}\hspace{1cm}
\includegraphics[angle=0,height=0.3\textwidth]{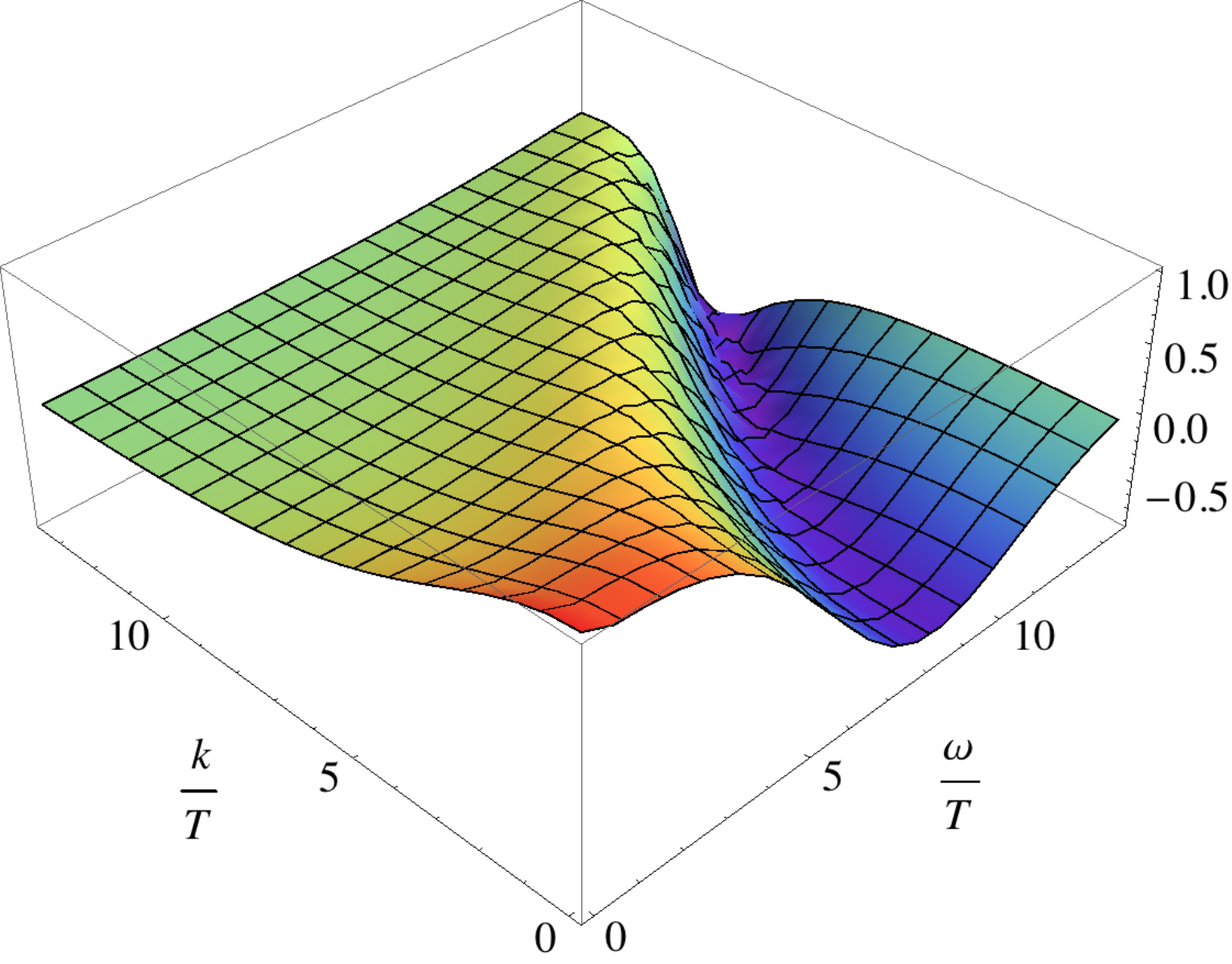}
\caption{Left plot shows the real part of $\sigma^\cV_5(\omega,k)$ and the right
plot is the real  part of  $\sigma^\cB_5(\omega,k)$. [Color online]}
\label{fig:sigma5Vwp}
\end{figure}
\end{center}

The frequency dependence of the magnetic conductivities in a
holographic model was studied first in \cite{Yee:2009vw} but the mixed
gauge-gravitational anomaly was not included. For that reason and for
completeness we also show in figure~\ref{fig:sigmaBStrong} these
conductivities. Notice that at the temperatures considered the
conductivities are not affected by a change in temperature (when they
are plotted as a function of $\omega/T$) and the position of the peak
of the imaginary part is at $\omega\sim 5T$ as in the weakly coupled
case for high temperatures, in agreement with the result for
$\sigma^\cB_e$ in~\cite{Yee:2009vw}. The only difference in the
frequency dependence introduced by the mixed anomaly with respect to
the former reference is the small jump in the conductivity at small
frequency (see green curves in Fig.
\ref{fig:sigmaBStrong})\footnote{We compare only with this curves
  because in~\cite{Yee:2009vw} the author studied only one anomalous
  $U(1)$. In consequence $\sigma_5^{\cB_5}$ is the conductivity to
  compare with (see also \cite{Amado:2011zx}). }. A difference with
Sec.~\ref{sec:weak_coupling} is also remarkable: the frequency
dependence of $\sigma^{\cB_5}_5(\omega)$ is slightly different from
$\sigma^\cB_e(\omega)$.  There is another qualitative similarity with
the model of Sec. \ref{sec:weak_coupling}: for both theories the
frequency in which the magnetic conductivities vanish is $\omega\sim
15T$. The system becomes ``insulator" if the frequency of the magnetic
field is higher than this specific value.
Then we solved the system for very low temperatures to study the zero
temperature behavior. In figure \ref{fig:sigmaBStronglow} we can see in the 
magnetic conductivities a jump at zero frequency as in the weakly coupled case, 
plus a resonance at $\omega\neq 0$. 
Another  similar feature with weak coupling is the plateau at small frequencies
in the
chiral separation effect.
To finish we study the frequency and momentum dependence of the conductivities.
The 
strongly coupled regime does not show a qualitative difference with respect to
the weakly 
coupled case in the regime of interest. We show in figure~\ref{fig:sigma5Vwp}
the
conductivities, $\Re[\sigma_5^\cV(\omega,k)]$ and $\Re[\sigma_5^\cB(\omega,k)]$.
The vortical 
conductivities in both models, the  weakly coupled and strongly coupled one,
require 
inhomogeneities in the vortex in order for a current being produced. The
phenomenological 
implications for the chiral vortical effect in the quark gluon plasma  will be
discussed in the section \ref{sec:discussions}. 

\section{Two point functions in hydrodynamics}
\label{sec:hydro}
To have a better understanding of the results obtained in the previous sections,
we will compute the form predicted by hydrodynamics of the two point functions
of interest. To do so we start with the first order constitutive relations for a
fluid with an anomalous $U(1)$.~\footnote{For simplicity and without loss of
generality we will consider a single anomalous $U(1)$. The extension to the symmetry
group $U_V(1)\times U_A(1)$ is straightforward.}~\footnote{These are the
constitutive
relations for the covariant current. See the Appendix for a discussion of the
difference 
between the consistent and covariant definitions of currents.}

\begin{eqnarray}
\nonumber T^{\mu\nu} &=& (\epsilon + P) u^\mu u^\nu + P g^{\mu\nu} -
\eta\sigma^{\mu\nu} + \sigma_\epsilon^\cB(B^\mu u^\nu + B^\nu u^\mu )  +
\sigma_\epsilon^\cV(\omega^\mu u^\nu + \omega^\nu u^\mu ) \,, \\
J^\mu  &=& \rho u^\mu +  \sigma^\cB B^\mu  + \sigma^\cV\omega^\mu\,.
\end{eqnarray} 
As the anomalous transport is also present in equilibrium, we also assume
electro-chemical equilibrium. This assumption allows us to remove the
thermoelectric terms. Apart from the constitutive relations we need the energy
conservation
\begin{equation}
\nabla_\mu T^{\mu\nu} = F^{\nu\mu}J_\mu  \,,
\end{equation}
and solve for the velocities which are the unknown variables in the system. To
do so, we will consider small fluctuations for the background fields $(h,A)$, $g_{\mu\nu}=\eta_{\mu\nu}+ h_{\mu\nu}(t,x)$ and $ A_\mu(t,x)$ and we will expand the expressions up to first order in them. These fluctuations
will take the fluid away from equilibrium, so that the new fluid velocity can be
written as $u^\mu = (1, v^i(t,x))$, where $v^i$ will also be small. In
particular to study the shear
sector it is necessary to switch on only $h_{ty}(t,x),$ $h_{tz}(t,x),$
$h_{xy}(t,x),$ $h_{xz}(t,x)$ and $A_y(t,x),$ $A_z(t,x)$. After plugging all
these ingredients in the constitutive relations and Fourier transforming them,
we end up with
\begin{eqnarray}
T^{t i} &=& (\epsilon + P)v_i + Ph_{ti} - i k \epsilon_{ij}
\sigma_\epsilon^\cV(v_j+h_{tj}) - i k\epsilon_{ij} \sigma_\epsilon^\cB A_j  \,, \\
T^{x i} &=& - Ph_{xi} +i\eta(\omega h_{xi} -k v_i) \,,  \\
J^i &=& \rho v_i - i k\epsilon_{ij} \sigma^\cV(v_j + h_{tj}) - i k\epsilon_{ij}
\sigma^\cB A_j \,,
\end{eqnarray}
where $i,j=y,z$ and $\epsilon_{ij}$ is the antisymmetric symbol. The
conservation law is
\begin{equation}
-\epsilon_{ij}\sigma_\epsilon^\cB \omega k A_j  + (-i\omega(\epsilon+P) + \eta
k^2)v_i  -i\omega (P + \epsilon)h_{ti} - \epsilon_{ij}\sigma_\epsilon^\cV\omega k
(h_{tj}+v_j)  - \eta\omega k h_{xi} - i\omega A_i\rho = 0 \,.
\end{equation}
Using these equations one can solve for the velocities. Next we plug the solutions into the constitutive relations and use linear response to relate these
expressions with the two point functions of interest. Using the scaling limit
$(\omega,k) \rightarrow (z^2 \omega, z k)$ for $z\ll 1$  as appropriate for isolating the diffusion pole in the shear channel, we arrive at the
correlators
\begin{eqnarray}
\langle T^{ti}T^{tj} \rangle &=& -i\epsilon_{ij} k \sigma_\epsilon^\cV \frac{-D^2 k^4}{(\omega+i D k^2)^2} \,, \label{eq:TThydro} \\
\langle T^{ti}J^{j} \rangle &=& -i\epsilon_{ij}k \left(\sigma_\epsilon^\cB -\frac{\rho \sigma_\epsilon^\cV}{\epsilon+P}\frac{\omega}{\omega+iDk^2} \right) \frac{iD k^2}{\omega + i D k^2}   \,, \label{eq:TJhydro} \\
\langle J^{i}T^{tj} \rangle&=&  -i\epsilon_{ij}k \left(\sigma^\cV -\frac{\rho\sigma_\epsilon^\cV}{\epsilon+P}\frac{\omega}{\omega+iDk^2} \right) \frac{iD k^2}{\omega + i D k^2}   \,, \label{eq:JThydro} \\
\langle J^i J^j \rangle &=& -i\epsilon_{ij}k\left( \sigma^\cB - \frac{\rho}{\epsilon+P}\left[ 2\sigma^\cV - \frac{\rho\sigma_\epsilon^\cV}{\epsilon+P}\frac{\omega}{\omega + i D k^2} \right]\frac{\omega}{\omega+iD k^2}\right)  \,, \label{eq:JJhydro}
\end{eqnarray}
where the shear diffusion constant is defined as $D=\eta/(\epsilon+P)$. The mixed correlators
$\langle J T \rangle$ and $\langle T J \rangle$ are exactly the same
because $\sigma_\epsilon^\cB=\sigma^\cV$.  We also note that all three conductivities associated with 
correlators containing the energy current $T^{ti}$ vanish in the limit
$k \rightarrow 0$ at finite frequency $\omega$. This is just the same
behavior we have already observed in our explicit weak and strong
coupling calculations.
From these expressions we can compute the position of the maxima in the real
part of the correlators by just solving the equations
\begin{eqnarray}
\frac{d}{d\omega}\Re\langle T^{ti}T^{tj} \rangle\left|_{\omega_M}\right. &=&0 \,,\\
\frac{d}{d\omega}\Re\langle T^{ti}J^{j} \rangle\left|_{\omega_M}\right. &=&0\,, \\
\frac{d}{d\omega}\Re\langle J^{i}J^{j} \rangle\left|_{\omega_M}\right. &=&0\,.
\end{eqnarray}
The real parts are related to the imaginary parts of the conductivities, cf.
Eqs.~(\ref{eq:Kubo1}) and (\ref{eq:Kubo2}). These equations have the solutions
\begin{eqnarray}
\omega_M &=& \pm \frac{1}{\sqrt{3}}Dk^2 \qquad \textrm{for}\qquad \langle T^{ti}T^{tj} \rangle\, ,  \label{eq:omegaMTT}\\
\omega_M &\simeq& \pm  Dk^2\qquad\quad\;\;\, \textrm{for} \qquad \langle T^{ti}J^{j}\rangle \,, \quad  \langle J^{i}T^{tj}\rangle \quad \textrm{and} \quad  \langle J^{i}J^{j}\rangle \,. \label{eq:omegaMTJ}
\end{eqnarray}

The results derived in this section for the vortical conductivities are plotted 
in figure~\ref{fig:sigmaVcompare1}.\footnote{In presence of two $U(1)$ the way to
 translate the results of this section is changing
 $\rho\rightarrow\rho_A$, $\sigma^{\cB,\cV} \rightarrow \sigma^{\cB_A,\cV}_A$ 
and $\sigma^\cB_\epsilon \rightarrow \sigma^{\cB_A}_\epsilon$ with $A=e,5$. 
This rule is more subtle for the correlators of the type $\langle J_e J_5\rangle$,
 but anyway the conclusion will not be modified in this case. } 
The holographic results shown in figure~\ref{fig:sigmaVcompare1} have the 
same behavior as the hydrodynamic computation\footnote{ Notice that the mixed correlators 
are made of two pieces, a leading part where the shear pole appears as a single pole, and a 
subleading (proportional to the ratio
between the charge density and the energy density) with the shear pole appearing as a double pole. 
However the correlator among two energy momentum tensor only has the contribution of the double pole.
In consequence this correlator could be more sensitive to the higher order corrections we have neglected
 in this computation. That could be the reason for the stronger deviations between holography and hydrodynamics 
we see in $\sigma_\epsilon^{\mathcal V}$ as compared to the other conductivities, see figure \ref{fig:sigmaVcompare1}.}. In particular, the position of the maxima 
in the imaginary part of the conductivities agrees quite well with Eqs.~(\ref{eq:omegaMTT}) and 
(\ref{eq:omegaMTJ}). Actually  even in the magnetic conductivities
we can see at small enough frequency the effect of the presence of the diffusion mode in consistency with  Eqs.~(\ref{eq:JJhydro}) and (\ref{eq:omegaMTJ}),  see figure \ref{fig:sigmaBStrong}.

\section{Ward Identities}
\label{sec:Ward}
In this section we will show that the vanishing of the chiral vortical
conductivities for non zero frequency follows from
energy-momentum conservation.
We therefore study what we can learn from the Ward identities for
diffeomorphisms. In particular we want to study the correlators 
that correspond to the Kubo
formulas for the chiral vortical conductivities and the chiral magnetic
conductivity for the energy current.

We start with the form of the transformations on metric and gauge field
\begin{align}
 \delta g_{\mu\nu} &= - \nabla_\mu \epsilon_\nu - \nabla_\nu \epsilon_\mu \,\\
 \delta A_\mu &= -\partial_\mu(\epsilon^\nu A_\nu) - \epsilon^\nu F_{\mu\nu}\,,
\end{align}
under an infinitesimal diffeomorphism $x^\mu \rightarrow x^\mu + \epsilon^\mu$.
We assume that the Green functions are obtained from an effective action
$W[g_{\mu\nu},A_\mu]$ such that under
the diffeomorphism
\begin{equation}
 \delta W = \int d^4x \left( \frac{\delta W}{\delta g_{\mu\nu}(x)}\delta
g_{\mu\nu}(x) + \frac{\delta W}{\delta A_\mu(x)}\delta A_\mu(x) 
\right) =0\,.
\end{equation}
We have assumed that all mixed gauge-gravitational anomalies are shifted into
the 
axial current via a suitable renormalization scheme.
Since $\epsilon^\mu(x)$ is an arbitrary vector we can derive the local
diffeomorphism Ward identity 
\begin{equation}\label{eq:WImaster}
 \nabla_\mu\left( \frac{2}{\sqrt{-g}} \frac{\delta W}{\delta
g_{\mu\nu}(x)}\right) + 
 \nabla_\mu\left( \frac{1}{\sqrt{-g}} \frac{\delta W}{\delta A_\mu(x)}\right)
A^\nu(x) + 
 \frac{1}{\sqrt{-g}} \frac{\delta W}{\delta A_\mu(x)} F_\mu\,^\nu(x) = 0\,.
\end{equation}
We note that the energy-momentum tensor and the current are
\begin{align}
 T^{\mu\nu} &= \frac{2}{\sqrt{-g}} \frac{\delta W}{\delta g_{\mu\nu}(x)} \,\\
 J^\mu(x) &= \frac{1}{\sqrt{-g}} \frac{\delta W}{\delta A_\mu(x)}\,.
\end{align}
In the flat space limit we can write $ T^{\mu\nu} = 2 \delta W /
\delta h_{\mu\nu}$ for $g_{\mu\nu} = \eta_{\mu\nu} + h_{\mu\nu}$.
We can obtain the wanted Ward identities by differentiating Eq.
(\ref{eq:WImaster}) with respect to the sources $g_{\mu\nu}(y)$ and
$A_\mu(y)$. We have assumed here that the metric has Euclidean signature.
To obtain expressions for retarded Green's functions in the Minkowski
signature we need to analytically continue the metric and the frequency.
The analytic continuation in the metric implies that all Euclidean timelike
indices on operators obey $T^{\tau\mu} \rightarrow i T^{0\mu}$. The
Euclidean  frequency is analytically continued in the standard way $i\omega_n =
\omega+i\epsilon$. 

\subsubsection{Chiral vortical conductivity in energy current}
Let us start with another differentiation with respect to $g_{\mu\nu}(y)$. 
Since we only want Ward identities in the absence of external sources we set
$A_\mu =0$ from the outset. 
We observe 
\begin{align}
 \Gamma^\mu_{\mu\rho} & = \frac 1 2 g^{\mu\lambda} \left( \partial_\mu
g_{\lambda\rho} + \partial_\rho g_{\lambda\mu}-
\partial_\lambda g_{\mu\rho}\right) = \frac{1}{2} g^{\mu\lambda} \partial_\rho
g_{\mu\lambda} \,\\
 \partial_\rho \frac{1}{\sqrt{-g}} &= -\frac 1 2 g^{\mu\lambda}\partial_\rho
g_{\mu\lambda} \,.
\end{align}
This leads to the simplified identity
\begin{equation}\label{eq;simplified1}
 \partial_\mu \left( \frac{\delta W}{\delta g_{\mu\nu}(x)}\right) +
\Gamma^\nu_{\mu\rho} \frac{\delta W}{\delta g_{\mu\rho}(x)} =0\,.
\end{equation}
Since we need to differentiate only once with respect to the external metric it
is sufficient to use the linearized background
metric $g_{\mu\nu} = \eta_{\mu\nu} + h_{\mu\nu}$ in order to compute the
Christoffel symbol
\begin{equation}
 \Gamma^\mu_{\nu\lambda} = \frac 1 2 \eta^{\mu\sigma}\left( \partial_\nu
h_{\sigma\lambda} + \partial_\lambda h_{\sigma\nu}
 - \partial_\sigma h_{\nu\lambda}  \right) \,.
\end{equation}
Then we get the Ward identity
\begin{equation}
 \partial_\mu^x \Pi^{\kappa\lambda,\mu\nu}(y,x) +
\eta^{\nu\kappa}\partial_\mu^x\delta(x-y) T^{\mu\lambda}(x) + 
 \eta^{\nu\lambda}\partial_\mu^x\delta(x-y) T^{\mu\kappa}(x) - 
 \eta^{\mu\nu}\partial_\mu^x\delta(x-y) T^{\kappa\lambda}(x) =0\,,
\end{equation}
where we also used the definition
\begin{equation}
  \Pi^{\lambda\kappa,\mu\nu}(y,x) = 4\left.\frac{\delta^2 W}{\delta
g_{\lambda\kappa}(y) \delta g_{\mu\nu}(x)}\right|_{g_{\mu\nu}=\eta_{\mu\nu}}
\end{equation}
This definition includes the seagull term, so that 
\begin{equation}\Pi^{\lambda\kappa,\mu\nu}(y,x) = \left\langle
T^{\lambda\kappa}(y)
T^{\mu\nu}(x)\right\rangle + \mathrm{seagull}\,.
\end{equation}

We assume translational invariance such that
$\langle T^{\mu\nu}(x) \rangle = \langle T^{\mu\nu}(0) \rangle = T_0^{\mu\nu}$.
The Fourier transformed Ward identity is now
\begin{equation}\label{eq:WI_momentumspace}
k_\mu\left( \tilde \Pi^{\lambda\kappa,\mu\nu}(k) + \eta^{\nu\lambda}
T_0^{\mu\kappa} +
\eta^{\kappa\nu} T_0^{\mu\lambda} - \eta^{\mu\nu} T_0^{\kappa\lambda} \right) =
0\,.
\end{equation}
To be able to say something about the chiral vortical conductivity in
the energy
current we evaluate this Ward identity in the polarization $k_\mu =
(k_0,0,0,k_z)$, $\kappa=x$,
$\nu=y$ and $\lambda=\tau$ with $\tau$ being Euclidean time.   In any case,
none of the
contact terms in (\ref{eq:WI_momentumspace}) contributes for our choice of
polarization.
\begin{equation}
k_\tau \tilde\Pi^{\tau x, \tau y} + k_z \tilde\Pi^{\tau x, z y} =0  \, .
\end{equation}
Now we write $k_\tau = \omega_n$ and also analytically continue the $\tau$
indices in the
correlators and arrive at
\begin{equation}
-\omega_n \tilde\Pi^{0 x, 0 y} +  i k_z \tilde\Pi^{0 x, z y} =0 \, .
\end{equation}
Finally we get the Ward identity for the retarded Green's function by setting
$i\omega_n = \omega +i \epsilon$ such that
\begin{equation}
i (\omega + i \epsilon) \frac{\tilde\Pi^{0 x, 0 y}(i\omega_n =
\omega+i\epsilon,k_z)}{ik_z} =  
-  \tilde\Pi^{0 x, z y} (i\omega_n = \omega+i\epsilon,k_z)\,.
\end{equation}
Now we take the limit $k_z\rightarrow 0$. On the left hand side we
just obtain the frequency dependent chiral vortical conductivity
multiplied with the frequency.
The correlator on the right hand side is evaluated at
zero momentum. This correlator can be further constrained via
rotational symmetry.  Under a rotation by $\pi/2$ along the $x$-axis
$\tilde\Pi^{0 x, z y}(\omega,\vec{k}=0)$ transforms as a symmetric
tensor and this implies $\tilde\Pi^{0 x, z y}(\omega,0) =
-\tilde\Pi^{0 x, y z}(\omega,0) $.
Since the correlator is symmetric in the $y,z$ indices this implies
that it must vanish $\Pi^{0x,yz}(\omega,0)=0$.
 Therefore we find for the chiral vortical conductivity
\begin{equation}\label{eq:WIsigmaVepsilon}
\omega\, \sigma^\cV_\epsilon (\omega) = 0 \,.
\end{equation}

\subsection{Chiral vortical conductivity in charge current}
Again we start from the Ward identity (\ref{eq:WImaster}). Now we want to
differentiate with
respect to the external gauge field. Therefore we can directly set $g_{\mu\nu} =
\eta_{\mu\nu}$
in (\ref{eq:WImaster}). Differentiating with respect to $A_\lambda(y)$ and then
setting $A_\mu=0$
and doing the Fourier transform as before gives
\begin{equation}\label{eq:WI_momentumspace2}
k_\mu\left( \tilde G^{\lambda,\mu\nu}(k) + J^\mu_0\eta^{\lambda\nu} -
J^\lambda_0\eta^{\mu\nu}\right)=0\,,
\end{equation}
where
\begin{equation}
(2\pi)^4 G^{\lambda,\mu\nu}(k) = \langle J^\lambda(k) T^{\mu\nu}(-k)\rangle
\end{equation}
Going through the same steps as before we arrive at
\begin{equation}
i(\omega+i\epsilon) \left. 
\frac{\tilde G^{x,0y}(i\omega_n = \omega + i \epsilon,
k_z)}{ik_z}\right|_{k_z\rightarrow 0} =
 \tilde G^{x,zy}(i\omega_n = \omega + i \epsilon, \vec{k}=0)\,.
\end{equation}
Invariance under rotations around the $x$ axis implies as before $\tilde
G^{x,zy}(\omega,0)=0$ and
therefore
\begin{equation}\label{eq:WIsigmaV}
\omega\, \sigma^\cV(\omega) = 0\,.
\end{equation}

\subsection{Chiral magnetic conductivity in energy current}
Now we need to get the correlators in reversed order. In the Euclidean theory
this
is easy, since the functional derivatives with respect to the metric and the
gauge field
commute. Therefore we also have the Ward identity
\begin{equation}
\partial^x_\mu \langle T^{\mu y}(x) J^x(y) \rangle =0\,,
\end{equation}
and this leads then to 
\begin{equation}
i \omega \frac{\langle T^{0 y}(k) J^x(-k) \rangle}{ i k_z} = \langle T^{zy}(k)
J^x(-k) \rangle\,.
\end{equation}
Taking the limit $k_z\rightarrow 0$ and using the resulting invariance under
rotation along the 
$x$ axis gives 
\begin{equation}\label{eq:WIsigmaBepsilon}
\omega\, \sigma^\cB_\epsilon (\omega) = 0\,.
\end{equation}

The relations (\ref{eq:WIsigmaBepsilon}), (\ref{eq:WIsigmaV}) and
(\ref{eq:WIsigmaVepsilon}) indeed imply that the chiral vortical
conductivities and the chiral magnetic conductivity in the energy current vanish
for non-zero frequency as we have indeed
found in our explicit calculations at weak and strong coupling.

\section{Discussion}
\label{sec:discussions}

The main result in our study is the behavior of the chiral vortical
conductivities as a function of the frequency. The weak and strong coupling
analysis produce the same result for the conductivities associated with a
homogeneous and time dependent vortex, and it reads
\begin{equation}
\label{eq:sigmaV}
\sigma^\cV_A(\omega) = \sigma_{A,(0)}^\cV \left( \delta_{\omega,0} + i\pi
\omega\delta(\omega) \right) \,, \qquad A = e, 5, \epsilon \,.  
\end{equation}

In section \ref{sec:Ward} we understood that this behavior in the
vortical conductivities is a requirement of energy-momentum
conservation. That is the reason why the conductivities for free
fermions and for the strongly coupled model show the same
non-analyticity. Therefore, for any theory with a conserved stress
energy tensor the vortical conductivities at zero momentum must be of
the form of Eq.  (\ref{eq:sigmaV}). However, the magnetic
conductivities in the currents are not subject to this constraint
because they are computed via two point functions of charged
currents. Hence the frequency dependence of the magnetic
conductivities will be model dependent.  The non-commutativity of the
limits $\omega\rightarrow 0$ and $k\rightarrow 0$ in the magnetic
conductivities (of the currents) is still induced by the mixing
with the shear channel but it is of quite different nature and does
not lead to the behavior (\ref{eq:sigmaV}).\footnote{Recently
the non-commutativity of these limits for the magnetic conductivities
have been investigated in a weakly coupled limit in~\cite{Satow:2014lva}.
This was done however without taking into account the coupling to the 
energy-momentum tensor and consequently commuting limits were found.}

To better understand the meaning of Eq. (\ref{eq:sigmaV}) and the
response pattern in real time
we consider a test body initially at rest which we start to rotate
with constant angular velocity $\Omega_k$ at time $t=0$ such that the
driving force is $\Omega_k \Theta(t)$ for a selected wave-number
$k$. We use the hydrodynamic approximation (\ref{eq:JThydro}) for the
response function.  The real time response in the current is then
given by the Fourier transform

\begin{align}\label{eq:realtimeresponse}
J(t) & = \Omega_k \int \frac{d\omega}{2\pi} \e^{-i\omega t} \left(  \sigma_0^\cV \frac{ i D k^2}{\omega+i Dk^2}
- \frac{\rho \sigma^\cV_\epsilon}{\epsilon+P} \frac{ \omega i D k^2}{(\omega+iDk^2)^2} 
\right) \frac{i}{\omega+i\epsilon} =
\theta(t) \Omega_k \left[ \sigma_0^\cV(1 - e^{- D k^2 t}) - \frac{\rho \sigma^\cV_\epsilon}{\epsilon+P} Dk^2 t e^{-Dk^2t}\right]\,.
\end{align}

The non-analytic behavior is now exhibited by the non-commutativity of the
limits $t\rightarrow \infty$ and $k\rightarrow 0$ (for simplicity we assume
$\Omega_k$ to be finite in the limit $k\rightarrow 0$).
If we first take $t$ to infinity we end up with the equilibrium response
determined by the value of $\sigma_0^\cV$. On the other
hand, if we take $k$ to zero first, we find that there is actually no response at
any finite value of $t$. In most physical situations
the wave number is limited effectively by the inverse of the linear size of the
system, which provides an infrared cutoff. Of course,
the lifetime of the system should be long enough for the exponential in
(\ref{eq:realtimeresponse}) to decay. It is a tempting exercise
to insert some typical numbers for the strongly coupled quark gluon plasma. We
note that the momentum diffusion constant is given by
$D = \frac{\eta}{\varepsilon+p} \approx \frac{1}{4\pi T}$ where we assumed the
shear viscosity to obey $s = 4\pi\eta$ and neglected
the chemical potentials \footnote{This is the high temperature limit in which also the second term proportional to $\sigma_\epsilon^\cV$ in (\ref{eq:realtimeresponse}) gives only small corrections. More precisely we find that $\lim_{ T\rightarrow \infty} \left(\frac{\rho\sigma^\cV_\epsilon }{(\epsilon+P) \sigma_e^\cV}\right) = 1/6$. }.  Using as cutoff the typical size of a fireball created
in heavy ion collisions $L \approx 10 \,\mathrm{fm}$ and defining the decay time
as $t_c= 1/Dk^2$ we find 
therefore  $t_c \approx 4 \pi T L^2$. Putting in the units $T=
350\, \mathrm{MeV}$, $L=10\, \mathrm{fm}$ and the 
conversion factor $\hbar c = 197\, \mathrm{MeV\, fm}$ we obtain $ t_c \sim
2200\,\mathrm{fm/c}$. Since the lifetime of the
quark gluon plasma is limited to $\tau \sim 10 \mathrm{fm/c}$ this means that
there is essentially no response of such a  
droplet of strongly coupled quark gluon plasma to a forced rotation on such a
short time scale  compared with the size of the system ($\tau \ll TL^2$).
This very crude estimate should of course not be taken too seriously. The
physical situation in a heavy ion collision is much more complicated
and does not correspond to rotation driven by an external force to which our
response formulas apply. To really understand better
the role the chiral vortical effect plays in heavy ion collisions one needs to
set up an initial value problem and solve the hydrodynamic
evolution equations. This is a much more complicated problem and far beyond the
scope of this article. Nevertheless we think that our
considerations raise the question of how effective the chiral vortical effect
might be in systems of finite lifetime even if they
are large enough to be well modeled via hydrodynamics. Hopefully these questions
can be addressed via numerical methods in the near future.

\section{Acknowledgments}
We would like to thank Juan L. Ma\~nes for useful clarifications on
the seagull term, Ho-Ung Yee for discussions and especially Dam T. Son 
for a very useful discussion on the interpretation of our results.  
This work has been
supported in part by Plan Nacional de Altas Energ\'{\i}as
(FPA2011-25948 and FPA2012-32828), Spanish MICINN Consolider-Ingenio
2010 Program CPAN (CSD2007-00042), Comunidad de Madrid HEP-HACOS
S2009/ESP-1473. Also by European Union's Seventh Framework Programme
under grant agreements (FP7-REGPOT-2012-2013-1) no 316165,
PIF-GA-2011-300984, the EU program ``Thales'' and ``HERAKLEITOS II''
ESF/NSRF 2007-2013 and was also co-financed by the European Union
(European Social Fund, ESF) and Greek national funds through the
Operational Program ``Education and Lifelong Learning'' of the
National Strategic Reference Framework (NSRF) under ``Funding of
proposals that have received a positive evaluation in the 3rd and 4th
Call of ERC Grant Schemes''. The authors acknowledge also the support
of the Spanish MINECO's “Centro de Excelencia Severo Ochoa” Programme
under grants SEV-2012-0234 and SEV-2012-0249. E.M. acknowledges the
warm hospitality and partial support from the Instituto de F\'{\i}sica
Te\'orica IFT-UAM/CSIC, where parts of this work were carried out. The
research of E.M. is supported by the Juan de la Cierva Program of the
Spanish MINECO.

\appendix*
\section{Sum over Matsubara frequencies}
\label{sec:matsubara}
In this Appendix we would like to discuss a subtle point on the
definition of the currents and the chemical potentials that appear in the
calculations. In particular we want to point out that there are two,
usually equivalent ways of calculating the sum over Matsubara
frequencies and to analytically continue to Lorentzian signature.

The textbook way of introducing the chemical potential is as follows. Consider a
system of fermions with
creation and annihilation operators $c^\dagger_k$ and $c_k$. At zero
temperature and finite density all
states up to a maximum energy are occupied. For free fermions the Fermi energy
is just the chemical potential
$\mu$. Let us label this state by $|\mu\rangle$. The creation and annihilation
operators corresponding
to momenta $k$ such that $\omega(k)<\mu$ acting on the state $|\mu\rangle$
change roles because of
the Pauli principle. Within that range of energies we have
\begin{align}
 c_k |\mu\rangle = |\mu-1\rangle\,,\\
 c^\dagger_k |\mu\rangle = 0\,.
\end{align}
The state $|\mu-1\rangle$ is the state in which the fermionic quantum of
momentum $k$ is missing (a hole state).
Therefore in this momentum range $c_k$ acts as creation operator (of holes) and
$c^\dagger_k$ as annihilation 
operator. This motivates us to introduce a new Hamiltonian that counts energy not
with respect to the normal
ordered vacuum but with respect to the finite density state $|\mu\rangle$. The
Hamiltonian measuring
energy with respect to the normal ordered vacuum is 
\begin{equation}
\label{eq:normalH}
 H_0 = \sum_k \omega(k) c^\dagger_k c_k
\end{equation}
whereas the Hamiltonian measuring energies with respect to the Fermi energy are 
$ \hat H = \sum_k [\omega(k)-\mu] c^\dagger_k c_k $
Therefore it is natural to define a new Hamiltonian 
\begin{equation}
\label{eq:grandH}
 \hat H = H_0 - \mu Q
\end{equation}
where $Q=\sum_k c^\dagger_k c_k$ is the (fermion) charge operator. The grand
canonical ensemble can now actually be understood
as the canonical ensemble of the system described by the Hamiltonian $\hat H$. 

On the other hand, the microscopic dynamics of the underlying physics is
unchanged
even in the state $|\mu\rangle$, and therefore
it is still described by the Hamiltonian $H_0$\footnote{Even if the state
$ |\mu\rangle $ is not necessarily an eigenstate of $H_0$.}.
This point of view can be expressed in a different way. One does not modify the
underlying Hamiltonian but rather modifies their
wave functions by demanding boundary conditions that reflect the presence of the
occupied states. At finite temperature the proper 
way to do this is to modify the boundary conditions for the fermionic fields
according to \cite{Landsman:1986uw, Evans:1995yz}
\begin{equation}
 \Psi(t-\frac{i}{T}) = - \exp(-q\mu/T) \Psi(t)\,.
\end{equation}
Formally this can be understood as a field redefinition $\Psi(t) \rightarrow
\exp(i q \mu t) \Psi(t)$. If the symmetry $Q$ is gauged
this can also be seen as a (non-proper) gauge transformation.

Let us now consider a typical sum over (fermionic) Matsubara frequencies
$\omega_n = i (2n+1)\pi T$ arising in one-loop calculations
at finite temperature. Let $f(z)$ be a meromorphic function  with poles on the
real axis. The summand is $f$ evaluated at the Matsubara
frequencies. At finite chemical potential the Matsubara frequencies are shifted
by $\mu$. We define therefore a deformation 
$f_\mu(z) = f(z+\mu)$.  
 
The sum over Matsubara frequencies is
\begin{equation}
 T \sum_n f(i\omega_n + \mu) = T \sum_n f_\mu(i\omega_n) = \frac 1 2
\oint_{\mathcal{C}_n} \frac{dz}{2\pi i} f_\mu(z) \tanh\left(\frac{z}{2T}\right)
\,,
\end{equation}
where $\mathcal{C}_n$ is the sum of contours that enclose the poles of the
hyperbolic tangent in counter clockwise fashion.

Using $\tanh(x/2) = 1 -2 n_f( x)= -1 +2 n_f(-x)$ with the Fermi-Dirac
distribution function $n_f(x) = [\exp(x)+1]^{-1}$ and deforming the contours
$\mathcal{C}_n$ to the contours $\mathcal{C}_\pm$ we can write
\begin{equation}
 T\sum_n f_\mu(i\omega_n) = \int_{-i \infty}^{+i \infty} \frac{dz}{2\pi i}
f_\mu(z)  -  
\int_{-i \infty+\epsilon}^{+i \infty+\epsilon} f_\mu(z) n_f\left(
\frac{z}{T}\right) +
\int_{-i \infty-\epsilon}^{+i \infty-\epsilon} f_\mu(z) n_f\left(
\frac{-z}{T}\right)
\end{equation}
Where we have assumed that there are no poles of $f_\mu(z)$ on the imaginary
axes. The second and third terms can be evaluated using
Cauchy's theorem by closing the contours with large half circles such that the
exponentials from the distribution functions suppress
the contributions from the half circles in the limit of infinite radius (see Figure 10 ). Finally
we can Wick rotate the first integral to real frequencies.
We need to take into account, however, that the Wick rotated contour implies that
the poles in $f_\mu(z)$ on the real axes have to be
circumvented with a particular $i\epsilon$ prescription such that poles on the
positive real axes lie below the contour and poles 
on the negative real axes lie above the contour. We arrive therefore at
\begin{equation}
 T\sum_n f_\mu(i\omega_n) = i \int_{-\infty}^{+\infty} \frac{dk_0}{2\pi}
f_\mu(k_0 + i \epsilon\, \mathrm{sgn}(k_0) ) +
\sum_k \mathrm{Res}\left( f_\mu(\hat z_k^+)\right) n_f\left(\frac{\hat z_k^+}{T}
\right) -
\sum_l \mathrm{Res}\left( f_\mu(\hat z_l^-)\right) n_f\left(\frac{-\hat
z_l^-}{T} \right)\,.
\end{equation}

\begin{center}
\begin{figure}[tbp]
\includegraphics[scale=1]{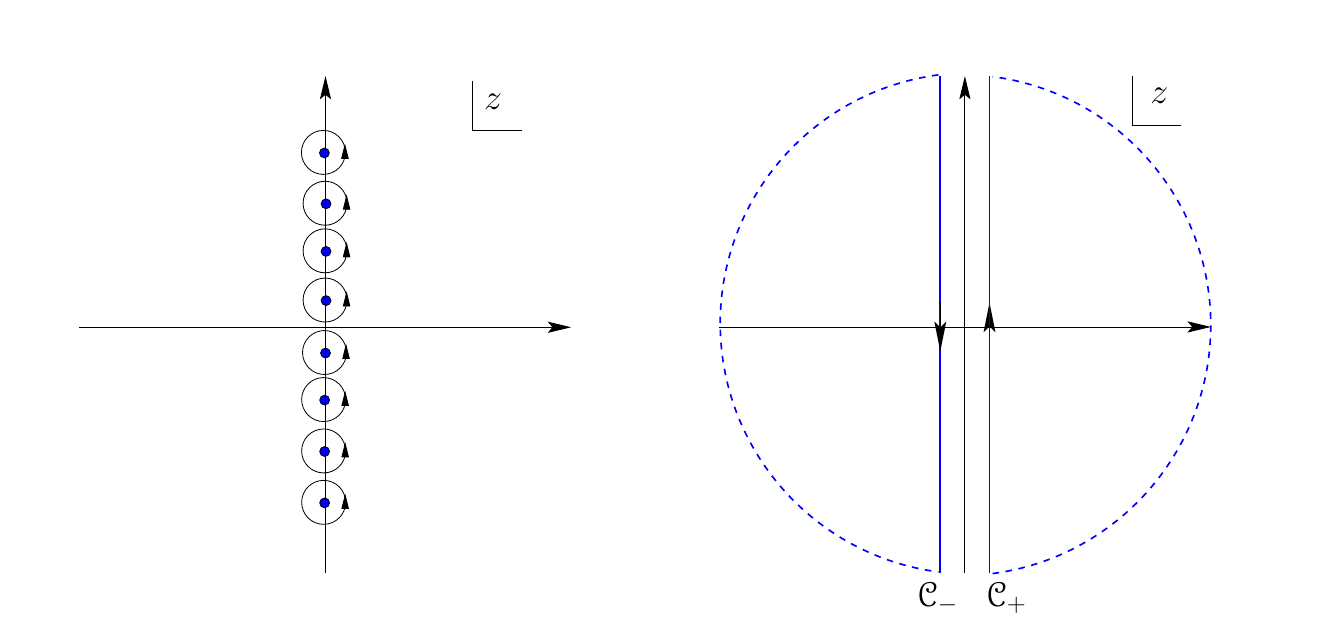}
\caption{The left part shows the contour encircling the Matsubara frequencies.
These contours can be deformed to
the ones shown in the right part.}
\label{fig:contours1}
\end{figure}
\end{center}

This formula can be interpreted in the following way. The first term is the
vacuum amplitude and the second and third terms are the
contributions from the thermally excited on-shell states, the second term
represents roughly speaking the particles (positive energy states
as measured by $\hat H$) 
and the third term represents the holes and anti-particles (negative energy states as
measured by $\hat H$). 
The first integral, is of course, in general divergent and well defined only with
a suitably chosen regularization. 
Note also that the second and third terms vanish at $T=0$. 

Were we to take the Hamiltonian $\hat H$ as the fundamental one, then the
first integral would indeed be the vacuum contour
corresponding to it. However, we are rather interested in
expressing the result in terms of the vacuum of the 
Hamiltonian $H_0$. Therefore it is necessary to express the Matsubara sum not in
terms of the function $f_\mu$ but rather
of $f$. We note that $\mathrm{Res}(f_\mu)(\hat z_k) = \mathrm{Res}(f)(z_k)$ with
$\hat z_k = z_k-\mu$, and therefore
\begin{equation}
 T\sum_n f_\mu(z) = i \int_{-\infty}^{+\infty} \frac{dk_0}{2\pi} f(k_0 + \mu + i
\epsilon\, \mathrm{sgn}(k_0) )  +
\sum_{z_k-\mu>0} \mathrm{Res}\left( f(z_k)\right) n_f\left(\frac{z_k-\mu}{T}
\right) -
\sum_{z_l-\mu<0} \mathrm{Res}\left( f(z_l)\right) n_f\left(\frac{\mu-z_l}{T}
\right)\,.
\end{equation}

Now we express the first contour in terms of the vacuum contour corresponding to 
the Hamiltonian $H_0$. This is the contour for which the
poles with $z<0$ lie above and the poles with $z>0$ lie below. The difference
between the vacuum contours corresponding to the Hamiltonians $\hat H$,
$H_0$ is a contour enclosing all poles between $z=0$ and $z=\mu$ (see Figure 11),  
\begin{equation}\label{eq:vacuumcontours}
i \int_{-\infty}^{+\infty} \frac{dk_0}{2\pi} f(k_0 + \mu + i \epsilon\,
\mathrm{sgn}(k_0) ) =  
i \int_{-\infty}^{+\infty} \frac{dk_0}{2\pi} f(k_0 + \mu + i \epsilon\,
\mathrm{sgn}(k_0+\mu) ) + \oint_{\mathcal{C}_\mu}\frac{dz}{2\pi i}f(z) \,. 
\end{equation}
Now we have succeeded in expressing the sum over Matsubara frequencies as a sum of
a vacuum integral, the finite density contribution  and
the thermal fluctuations.  Note that the poles of $f$ inside the contour
$\mathcal{C}_\mu$ are precisely those poles for which either
$z_k>0$ and $z_k-\mu<0$ for positive $\mu$ or $z_l<0$ and $z_l-\mu>0$ for
negative $\mu$. Using $n_f(-x) = 1-n_f(x)$ for these poles
in the thermal sums we can express the amplitude as
\begin{equation}\label{eq:matsubaraA}
 T\sum_n f_\mu(z) = i \int_{-\infty}^{+\infty} \frac{dk_0}{2\pi} f(k_0 + \mu + i
\epsilon\, \mathrm{sgn}(k_0+\mu) )  +
\sum_{z_k>0} \mathrm{Res}\left( f(z_k)\right) n_f\left(\frac{z_k-\mu}{T} \right)
-
\sum_{z_l<0} \mathrm{Res}\left( f(z_l)\right) n_f\left(\frac{\mu-z_l}{T}
\right)\,.
\end{equation}

\begin{center}
\begin{figure}[tbp]
\includegraphics[scale=1]{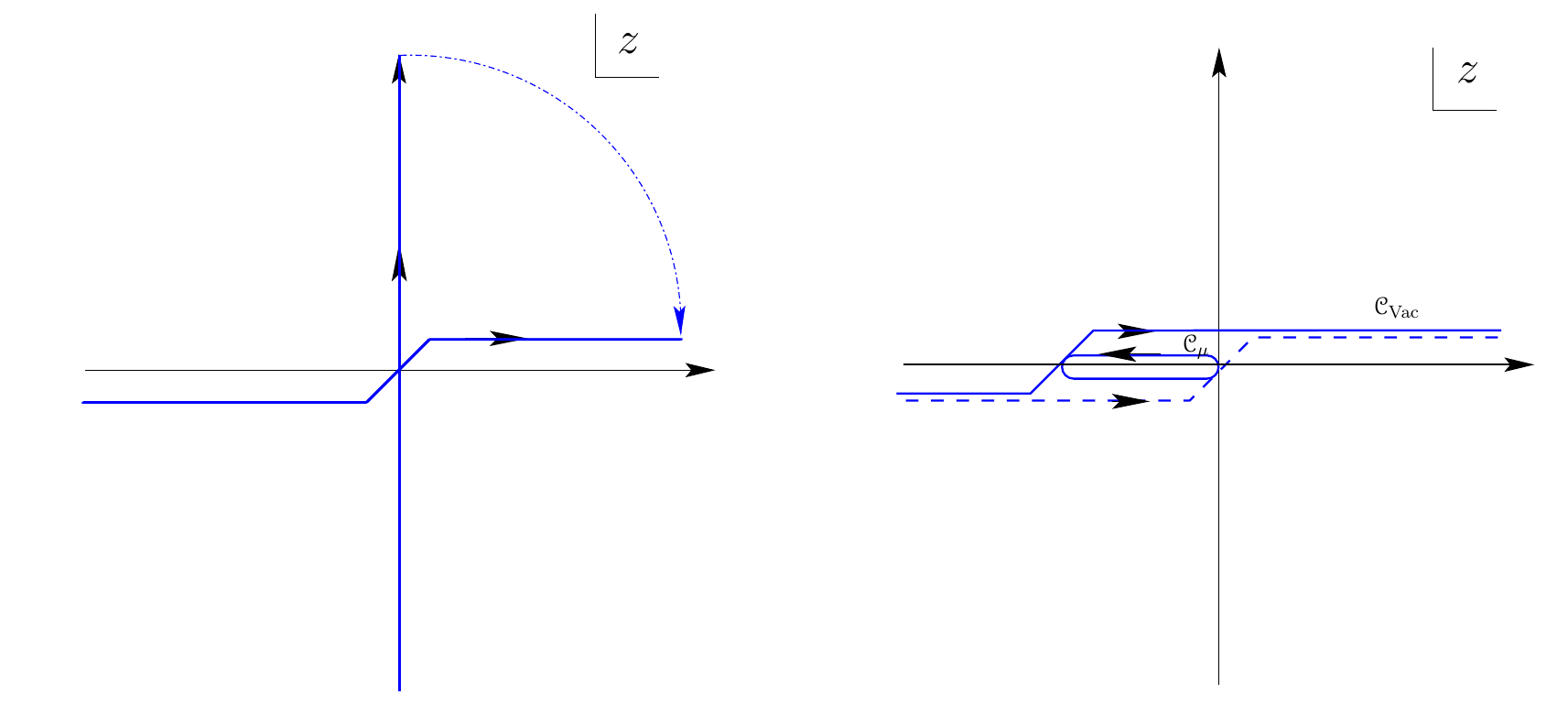}
\caption{The Euclidean vacuum contour can be Wick rotated to a Lorentzian one.
By another contour deformation is  written as the vacuum contour
of the normal ordered vacuum of the Hamiltonian $H_0$ (possibly in the gauge
$A_0=\mu$
plus a finite contour enclosing the poles corresponding to the occupied states
up to $k_0 = \mu$.}
\label{fig:contours2}
\end{figure}
\end{center}

The thermal sums over the residues are expressed now in terms of the particles
and
anti-particles according to the usual normal ordering
prescription, i.e. the Hamiltonian $H_0$. They do not vanish at $T=0$ since
they implicitly contain the finite density contribution.

Let us now use another way of defining the sum over Matsubara frequencies. We
could also write
\begin{equation}
 T \sum_n f(i\omega_n + \mu) = \frac{1}{2} \oint \frac{dz}{2\pi i} f(z)
\tanh\left( \frac{z-\mu}{2T}\right)\,.
\end{equation}
The only difference now is that the contours enclosing the poles of the
hyperbolic tangent all are centered at $\Re(z) = \mu$. 
Going through the same steps as before we now end up with
\begin{equation}
\label{eq:matsubaraB}
 T\sum_n f_\mu(z) = i \int_{-\infty}^{+\infty} \frac{dk_0}{2\pi} f(k_0  + i
\epsilon\, \mathrm{sgn}(k_0) )  +
\sum_{z_k>0} \mathrm{Res}\left( f(z_k)\right) n_f\left(\frac{z_k-\mu}{T} \right)
-
\sum_{z_l<0} \mathrm{Res}\left( f(z_l)\right) n_f\left(\frac{\mu-z_l}{T}
\right)\,.
\end{equation}

The difference between (\ref{eq:matsubaraA}) and (\ref{eq:matsubaraB}) is simply
a shift in the integration coordinate $z$ or $k_0$ respectively.
Such a shift in momentum can also be understood as a gauge-transformation, if we
assume for the moment that the symmetry generated by the charge $Q$ is
gauged. The shift in the integration variable $k_0$ is then equivalent to a
gauge transformation with a time dependent gauge parameter $\lambda = \mu t$.
This introduces the gauge field $A_0=\mu$ into the theory. Note, however, that
this
gauge transformation does not change anything in the contours we used
to define the Matsubara sums. The presence of the chemical potential is not
related to the presence of a gauge field $A_0=\mu$ but rather to the
way the poles of the function $f$ are circumvented in the complex energy
plane. 

Since (\ref{eq:matsubaraA}) and (\ref{eq:matsubaraB}) are related by the gauge
transformation $\lambda = \mu t$, we should ask now what happens
if the symmetry under consideration is anomalous, as this is the case of
interest in this paper. To be specific we consider a chemical potential
for the axial symmetry. It is well known that the axial anomaly for one Dirac
fermion takes the form
\begin{equation}\label{eq:axialanomaly}
 \delta_5 W_{\mathrm{eff}}[A,g] = \int
\sqrt{|g|}\lambda_5\epsilon^{\mu\nu\rho\lambda} \left[ \frac{1}{48\pi^2} \left(
F^5_{\mu\nu} F^5_{\rho\lambda} +
3 F_{\mu\nu} F_{\rho\lambda}  \right) + \frac{1}{384\pi^2}
R^\alpha\,_{\beta\mu\nu} R^\beta\,_{\alpha\rho\lambda}
\right]\,,
\end{equation}
where $\delta_5$ is an axial transformation with parameter $\lambda_5$. We have
introduced (so far non-dynamical) gauge fields for the axial, vector 
and diffeomorphism symmetry, $A^5_\mu$, $A_\mu$ and the metric $g_{\mu\nu}$. The
anomaly is actually the sum of the pure axial anomaly 
given by the terms with the axial field strength $F^5_{\mu\nu}$, the mixed
axial-vector anomaly and the mixed axial-gravitational anomaly. All anomalies
have been shifted into the axial current. This implies a specific regularization
scheme in evaluating the vacuum graph in (\ref{eq:matsubaraB}).

Let us see what are the consequences of using either (\ref{eq:matsubaraA}) or
(\ref{eq:matsubaraB}) for evaluating the Kubo formulas for anomalous
transport in the presence of an axial chemical potential $\mu_5$. The gauge
parameter relating the two is $\lambda_5 =  \mu_5 t$. If we define
the currents as the functional derivatives of the effective action, then the
anomaly (\ref{eq:axialanomaly}) will give new contributions beyond the
common terms captured by the thermal sums of the residues. We are only
interested here matching our results up to terms of first order in derivatives
on the background fields. It follows that at this order of derivatives the mixed
axial-gravitational anomaly will not contribute. We will further
specify to the case where only a magnetic field $B_i = -\frac 1 2
\epsilon^{0ijk} 
F_{jk}$ is present\footnote{We define $\epsilon_{0123}=+1$ therefore
$\epsilon^{0123}=-1$.}.

It is now straightforward to see that the contribution of
(\ref{eq:axialanomaly}),
which arises from the vacuum integral in formula (\ref{eq:matsubaraA}) is given
by
\begin{equation}\label{eq:notCME}
 \vec{J}_{\mathrm{vac}} = -\frac{\mu_5}{2\pi^2} \vec{B}\,.
\end{equation}
The perturbative evaluation of the vacuum graph in (\ref{eq:matsubaraA}) can be
done by expanding in $\mu_5$. At first order one encounters
a triangle diagram with $\mu_5$ sitting at one of the vertices. This
diagram has been explicitly evaluated in \cite{Gynther:2010ed} with the result
(\ref{eq:notCME}). 
This is precisely the negative of the chiral magnetic 
effect stemming from the finite $T,\mu_5$ contribution! Therefore the total
chiral magnetic
effect vanishes if formula (\ref{eq:matsubaraA}) is used for the axial chemical
potential.
This was first noticed 
in a holographic context in \cite{Rebhan:2009vc}.
Therefore one needs to distinguish between the axial chemical potential and
the axial background gauge field. Since axial gauge fields are absent on a
fundamental
level in nature we therefore advocate to use formula (\ref{eq:matsubaraB}).

In the presence of anomalies the current defined as the variation of the effective
action with respect to the gauge field is called the consistent
current (because the anomaly (\ref{eq:axialanomaly}) has to fulfill the
Wess-Zumino consistency condition). This current is not invariant under
the anomalous gauge transformation. But it is possible to define a current that
is invariant under all gauge transformations, anomalous and
non-anomalous ones by adding suitably chosen Chern-Simons currents. The axial
gauge variation of the electric current follows from 
(\ref{eq:axialanomaly}) as
\begin{equation}
 \delta_5 J^\mu = \frac{1}{4\pi^2} \epsilon^{\mu\nu\rho\lambda} \partial_\nu
\lambda_5 F_{\rho\lambda}\,.
\end{equation}
There is, of course, nothing wrong with this. Axial transformations are not a
symmetry and so there is no reason for the (consistent) current
to be invariant. However it is now easy to define an invariant current by adding
the Chern-Simons current
\begin{equation}
 J^\mu_{\mathrm{cov}} = J^\mu - \frac{1}{4\pi^2} \epsilon^{\mu\nu\rho\lambda}
A^5_\nu F_{\rho\lambda}\,.
\end{equation}
This current is invariant under the usual gauge transformations
and also under axial transformations.
The covariant current is not conserved, rather it fulfills 
\begin{equation}\label{eq:covanomaly}
 \partial_\mu J^\mu_{\mathrm{cov}} = -\frac{1}{8\pi^2}
\epsilon^{\mu\nu\rho\lambda} F^5_{\mu\nu} F_{\rho\lambda} \,.
\end{equation}
We also note that in the absence of an axial background gauge field the
covariant 
expectation values of the consistent currents coincide. One can also
introduce a covariant axial current via 
$J^\mu_{5,\mathrm{cov}} = J^\mu_5 - \frac{1}{12\pi^2}
\epsilon^{\mu\nu\rho\lambda}
A^5_\nu F_{\rho\lambda,5}$.

We intend now to give a physical interpretation of the consistent and covariant
currents. From equation (\ref{eq:matsubaraA}) we see that
the finite temperature and density contributions are given by well
defined expressions. Integrals over spatial momenta are regulated
by the presence of the Fermi-Dirac distributions. All contributions
to the current stemming from them are automatically invariant.
Therefore we identify the current produced by the physically moving charges,
i.e. the collective motion of the on-shell states, as the contributions to the
covariant
current. The covariant anomaly (\ref{eq:covanomaly}) states that in parallel
electric and magnetic fields of axial and vector type charged particles
are created out of the vacuum via spectral flow. 
More precisely, for every finite ultraviolet cutoff particles are flowing into
the physical region below that cutoff via the spectral
flow induced by the parallel electric and magnetic fields. 

The covariant current is, however, not the one that couples to the (vector) gauge
field $A_\mu$. This is by definition the consistent current
with exactly vanishing divergence. If we give dynamics to the vector field by
adding a Maxwell term to the effective action, it is the
consistent current that enters Maxwell's equation
\begin{equation}
 J^\nu = \partial_\mu F^{\mu\nu}\,,
\end{equation}
whereas the covariant current enters in a Chern-Simons modification of Maxwell's
equation
\begin{equation}\label{eq:maxcs}
 J^\nu_\mathrm{cov} = \partial_\mu F^{\mu\nu} -
\frac{1}{4\pi^2}\epsilon^{\nu\mu\rho\lambda}A^5_\mu F_{\rho\lambda}\,.
\end{equation}

Let us summarize now the response to a magnetic field in the covariant, the 
consistent and the energy currents.
We also keep now the notion of chemical potential and (axial) vector potential
apart but do include
the presence of constant $A^5_0$. Then
\begin{align}
 \vec J_{\mathrm{cons}} &=  \frac{\mu_5 - A^5_0}{2\pi^2}\vec{B}\,,
\label{eq:cmecons}\\
 \vec J_{\mathrm{cov}} &=  \frac{\mu_5 }{2\pi^2}\vec{B}\,, \label{eq:cmecovii}\\
 \vec{J}_\epsilon &= \frac{\mu \mu_5}{2\pi^2} \vec{B}\label{eq:cmeenergy}\,.
\end{align}

Note that the energy flow does {\em not} depend on the axial gauge field
background! In fact even upon using (\ref{eq:matsubaraA})
the Kubo formula for the energy flow does not depend on the presence of a
background axial gauge field because the 
perturbative expansion of the vacuum amplitude involves a triangle diagram of
the form $\langle T^{0i} J^j J_5^0\rangle A_0^5$.
This triangle diagram does not have an anomalous contribution and is therefore
insensitive to a constant $A_0^5$. 
So in the absence of $\mu_5$ there is no energy flow connected to the presence of
the consistent current (\ref{eq:cmecons}). 
Moreover, notice that if one chooses $A_0^5=\mu_5$ one gets a vanishing chiral
magnetic effect in the consistent current but
a non-vanishing result in the energy current!\footnote{In a the context of lattice field theory this has also been confirmed recently in
\cite{Buividovich:2013hza}.}\footnote{
Recently an instability in the photon propagator due to the chiral magnetic
effect has
been found in \cite{Kirilin:2013fqa}. Also in this context the distinction
between covariant and
consistent current seems important because it is only the consistent current
that enters Maxwell's equations.}.

A related aspect concerns the energy-momentum conservation in the presence of
external background
gauge fields. It follows from the Ward identity for diffeomorphisms
(\ref{eq:WImaster}) which can be written as
\begin{equation}
 \partial_\mu T^{\mu\nu} = F^{\nu\mu}J_\nu + F_5^{\nu\mu} J^5_\mu -
\frac{A^\nu_5}{48\pi^2} \epsilon^{\mu\lambda\rho\sigma} \left(
\ 3 F_{\mu\lambda} F_{\rho\sigma} + F^5_{\mu\lambda} F^5_{\rho\sigma}\right) =
0\,.
\end{equation}
Evaluating this in a frame where there is no energy current in the background of
vanishing axial field strength but presence of the chiral magnetic
current (\ref{eq:cmecons}) we find that the contributions from the axial
background field vanish and therefore
\begin{equation}
 \partial_t T^{00} = \frac{\mu_5- A_0^5}{2\pi^2} \vec{E}\vec{B} +
\frac{A_0^5}{2\pi^2} \vec{E}\vec{B} = \frac{\mu_5}{2\pi^2} \vec{E}\vec{B}\,.
\end{equation}
This seems natural since we have seen already in (\ref{eq:cmeenergy})
that the axial background field does not contribute to the energy 
transport. In fact the cancellation is more generally valid. The energy momentum
conservation is most conveniently expressed in terms of the
covariant currents as
\begin{equation}
 \partial_\mu T^{\mu\nu} = F^{\nu\mu} J_{\mu,\mathrm{cov}} + F_5^{\nu\mu}
J_{\mu,\mathrm{cov}}^5\,.
\end{equation}

One possible point of view is therefore that the Chern-Simons current
$J^\mu_{CS} = \epsilon^{\mu\nu\rho\lambda} A^5_\nu F_{\rho\lambda}$ is not a
genuine 
transport phenomenon.  Rather we might take it as the necessity to modify
Maxwell's equations
in a quantum vacuum of chiral fermions by adding a Chern-Simons current as in
(\ref{eq:maxcs})!  

\bibliographystyle{apsrev4-1} 
\bibliography{refs}

\end{document}